\documentstyle[preprint,aps,psfig]{revtex} 
\def\Journal#1#2#3#4{{#1} {\bf #2}, #3 (#4)}

\def\PRL{Phys. Rev. Lett.}
\def\PRD{{Phys. Rev.} D}

\def\CQG{{Class. Quant. Grav.}}
\def\JMP{{J. Math. Phys.}}

\def\be{\begin{equation}}
\def\ee{\end{equation}}
\def\bea{\begin{eqnarray}}
\def\eea{\end{eqnarray}}

\tighten
\begin{document}

\renewcommand{\thefootnote}{\fnsymbol{footnote}}
\def\scf{\setcounter{footnote}}
% example: \scf{1}\footnote{}

\title{An alternative approach to the galactic dark matter problem\\ }

\author{Ulises Nucamendi\footnote{e-mail: ung@star.cpes.susx.ac.uk} \\
\small {\it Centre for Theoretical Physics}\\
\small {\it University of Sussex}\\
\small {\it Brighton BN1 9QJ, Great Britain}\\
\vspace*{0.5cm}
Marcelo Salgado\footnote{e-mail: marcelo@nuclecu.unam.mx} and
Daniel Sudarsky\footnote{e-mail: sudarsky@nuclecu.unam.mx} \\
\small {\it Instituto de Ciencias Nucleares} \\
\small {\it Universidad Nacional Aut\'onoma de M\'exico} \\
\small {\it Apdo. Postal 70-543 M\'exico 04510 D.F, M\'exico}. }
\maketitle

\abstract{
We discuss  scenarios in which  the galactic dark matter in spiral
galaxies is described by a long range coherent field which
settles in a stationary configuration that might account
for the features of the galactic rotation curves.
 The simplest possibility is to consider  scalar fields, so we discuss
in particular, two mechanisms that would account for the settlement of the 
scalar field in a non-trivial configuration in the 
absence of a direct coupling of the field with ordinary matter:
topological defects, and spontaneous scalarization.}

 \vskip 1.5cm
\noindent PACS number(s): 11.27.+d, 04.40.-b, 98.62.Gq

\newpage
\newpage

\section{Introduction}
 It has been known for a long time that the motion of the stars and gases
around the center of most galaxies can not be explained in terms of the
luminous matter content of the galaxies, at least not within the context of
Newtonian gravity (see \cite{rubin} for a review). The standard view is
that there is in almost every galaxy
a large component of non-luminous matter (the gravitational dark matter) 
that forms an halo around the galaxy and that provides the additional
gravitational attraction needed to explain the ``rotation curves'' in terms of
standard gravitational theory. There are several proposals
for this  dark component, ranging from new exotic particles such as those
predicted by supersymmetry \cite{turner,primack,susy},
to other less exotic candidates such as massive neutrinos,
all collectively known
 as  WIMPs (weakly interacting massive particles) (see \cite{turner,kolb} for
a review), to the relative mundane
 idea of dark but ordinary bodies such as Jupiter-like objects
 collectively known as MACHOs (Massive Compact Halo Objects)\cite{carr}.
Searches for these types of objects have been made \cite{alcock}, and
although they report some findings, there doesn't seem to be
enough of these objects to account for the
galactic dynamics.
Moreover there are severe bounds on the amount of  baryonic matter in the
universe arising from big bang
nucleosynthesis and for some values of the Hubble constant those bounds
also imply that some of the galactic dark matter ought to be 
exotic\cite{peebles1,weinberg,copi}. Independently of this and 
despite their popularity, these 
type of models suffer from various problems and require surprising 
coincidences (see for example \cite{sellwood}).\\
\\
Another type of proposal, which is in some sense more radical, is based
on the idea that the gravitational theory would have to be modified when
dealing
with the scales associated with the motion of stars in galaxies
\cite{finzi,sanders}, in
particular the idea is embodied by the proposal of Milgrom 
\cite{milgrom}, that the laws of
motion are modified  when the accelerations involved are extremely small.
Unfortunately this scenario has not, as yet, been converted into a fully
relativistically invariant theory. Another type of model that has been
exposed is to
replace general relativity by a  higher order in curvature theory, which
in  some particular cases appears to
be obtaining encouraging results\cite{Manheim}. The problem with this
approach is that these types of theories
have in general problems of principle like for example the lack of a well
posed initial value formulation.
  Nevertheless 
such relatively radical proposals are still attractive, due in part to an
intrinsic problems of the more conservative approaches
 in explaining the generality and universality of the phenomena, namely
 the   fact that the amount of luminous matter seems to be such a good
indicator of the amount of the dark matter component \cite{persic}
and the fact that the
dark component happens always to distribute itself in such a way that
the resulting rotation curves (hereafter referred to as RC) 
are almost flat away from the  galactic centers \cite{rubin1}.  \\
\\
Thus, in contrast with the  former scenarios which would need to assume
not only the existence of the dark matter but also give some evolutionary
scenarios that result in the aforementioned universality in its
 distribution, the modified gravity scenario would naturally account
for such correlations without the need for additional assumptions. On the
other hand the former scenarios do not present any problem in lending
themselves to an acceptable theoretical formulation, compatible with
present theories of particle physics and general relativity.  \\
\\
The object of this article is to discuss a third type of scenario which
has some of the advantages of each scheme. The idea is to take dark
matter to be described not by a bunch of particles whose distribution
needs to be explained but by a coherent field which would settle in a
universal stationary configuration that would account for the generic
features of the RC. The simplest possibility is provided by
scalar fields, which would of course have to be very long ranged (i.e.
masses smaller than ${1 \over R_G}$ where $R_G$ is the radius of the largest
galaxy with flat RC). The basic problem is that there are
very severe experimental bounds for the direct coupling of such a
field with ordinary matter \cite{panov}, and in the absence
of such coupling the field will in general settle globally in the minimum of
the potential leading to a homogeneous configuration that will not produce
the desired effects. On the other hand, one could hope that, given the
likelihood of existence of large black holes at the center of most
galaxies, they would account for the nontrivial configuration of the
scalar fields. Unfortunately these kind of situation is largely
forbidden by the ``Black Hole No hair theorems'' for scalar fields 
\cite{MH,DS,JDB}. These limitations severely reduce the types of 
models one can consider, in particular, there are, known to these 
authors, only three
mechanisms that would account for the settlement of the scalar field in a
non-trivial configuration in the absence of a
direct coupling of the field with ordinary matter (or some other exotic
matter which we will not consider because of the incremental number of
hypothesis it involves): a)  boson-star like clumps,
   b) spontaneous
scalarization, and  c) topological defects. Other models that lack these 
features have been considered, for example in \cite{tonatiuh}. However, 
such models face two problems: first, they give rise 
to spherical configurations where the scalar field in consideration is 
singular ``at the center'', and second, the resulting scalar field 
potential needed to account for the flat RC depends explicitly on the 
value of the ``tangential velocity'' of stars at the flat region. 
That is to say, such a potential have to be adjusted differently for 
different galaxies. Needless to say that both problems clearly make 
those schemes unsuitable as models for the problem at hand.
\bigskip

Concerning the case ``a)'' mentioned above, it has been 
analyzed in \cite{peebles}.
Their analysis focuses on cosmological and evolutionary considerations
as well as the issues related to the conditions under which the assumption
of long range coherency of the
 scalar field is justified, rather than the universal features of the
galactic rotation curves.
We will deal here with the other two cases ``b)'' and ``c)''.  \\
\\
The scenario ``b)'', namely the spontaneous scalarization
(see Sec.V)\cite{Damour}, is in some sense simpler because it involves a
single scalar field in contrast of the various fields needed in the
simplest versions of topological defects (e.g., global monopoles). 
Here the mechanism that allows for the
nontrivial stationary configuration of the scalar field is connected  to a
non-minimal coupling of the scalar field to the curvature. This results in
the effective gravitational coupling becoming dependent on the scalar field.
The point is that such a coupling allows for the reduction of the total
energy of the configuration (in comparison with the
corresponding configuration with the
same baryon number and no scalar field)
 for which the scalar field deviates from
the trivial configuration by taking values that reduce the gravitational
coupling in the regions of high matter density \cite{SSN}. Thus the model
must incorporate from the onset the non-minimal coupling that seems to be
needed to account at the same time for the correlations in the
dark-luminous matter components (see \cite{NSS} and 
the discussion of the third scenario 
below). The disadvantage of this model, which is
in fact shared by the
first model (i.e., boson stars) \cite{Igor}, is precisely
the lack of resilience against black holes whose existence in most galaxies,
if confirmed, would seem to preclude, through the no-hair theorems
\cite{MH,DS,JDB},
the models based on this mechanism.  \\
\\
The scenario ``c)'' is exemplified by the model of
global  monopoles \cite{vilenkin1} which have the notable feature of naturally
leading to a $1/r^2$ energy
density behavior which would naively account for the flat rotation curves and
which upon taking the symmetry braking scale to be the GUT scale
would result in the correct order of magnitude for the galactic
dark matter. Unfortunately, upon further examination of the simplest model
severe problems arise, in particular the monopole configuration turns out
to be repulsive \cite{harari}, and moreover the configuration would be too
universal in the sense that it would be independent of the size of the
galaxy thus defeating the hope for the correlation  of dark to luminous matter
 over a range of galactic sizes. There is nevertheless hope to overcome
 these problems by the consideration of slightly more complicated models
\cite{NSS}. In that work  the simple monopole model was supplemented
by the introduction of a non-minimal coupling between the
scalar fields and curvature (see \cite{NSS} and Sec.VI). This resulted in the restoration of
gravitational attraction leading to regions of relatively flat rotation
curves and to the possibilities of the dark-luminous matter correlations
arising from the fact that in these models the scalar potential
$V(\Phi^a \Phi_a)$ (where $\Phi^a$ stands for a triplet of scalar fields that characterize the 
global monopole) is replaced with the effective scalar potential $V(\Phi^a\Phi_a)+ F(\Phi^a\Phi_a,R)$ 
(here $R$ stands for the scalar curvature of the space-time metric)
whose minima depend on the amount of matter present trough  the effect
of the latter on $R$.  The global monopole model has the additional
advantage of resilience against the formation of black holes in the galactic
centers, since their topological charge makes them immune to the devastating
limitations imposed by the no hair theorems. \\
\\
Despite the promising
features of the model ``c)'', our intention in the present work is to take a 
step backwards and look at the problem from a more general point of view
before embarking in the methodical study of a particular type of model. \\
\\
The article is organized as follows: in Section II, we analyze the
generic form of the rotation curves of galaxies in a general relativistic
context. In section III, we comment on the  Newtonian approximation and 
on the embedding of the galaxy in the large scale space-time.
In Section IV, we discuss the additional information that can be
obtained about the metric from other considerations, specifically the
deflection of light by the galaxy.
Section V reviews the spontaneous scalarization
scenarios. In Section VI, we
review the non-minimally coupled global monopole model and discuss its
shortcomings. Finally, in Section VII we offer a discussion and analyze the
directions for further developments.
\bigskip

\section{Rotation curves of galaxies and frequency shifts}

The rotation curves (RC) provides the most direct method
of analyzing the gravitational field inside  a spiral galaxy.
RC have been determined for a great amount of spiral galaxies
\cite{persic,rubin1}. They are obtained
by measuring the frequency shifts of light emitted from stars
and from the 21 cm radiation from neutral gas clouds.

In fact, since (apart from the central regions) 
the ``tangential velocity'' of rotation  $v$ remains
 approximately constant up to distances far beyond
the  luminous radius of  these galaxies, a naive Newtonian analysis leads
to the conclusion
that the energy density decreases with the distance as
$r^{-2}$ and therefore that the mass of galaxies 
increases as $m(r) \approx r$. On the other hand, one could naturally question
whether these large mass ought not to result into an important
gravitational redshift.
We will  carry  our analysis in a general relativistic setting and will
see in the following
sections  that with standard assumptions about the matter content of the galaxy, 
the behavior of the RC 
indicates that the spacetime is not in general  described by the standard form
$ds^2= -(1+2\Phi) dt^2 + (1+2\Phi)^{-1} dr^2 +r^2 d\Omega$
 as can be  initially thought (see for example \cite{peebles}). \\
\\
In order to  analyze the problem we will
focus  directly on  what it is  observed because only then will we be able
to discuss models that do not
let themselves to a Newtonian based inferences. This is an important point
since the
lack of understanding of it leads to erroneous conclusions\cite{Edery}. \\
\\
The observations of stars and gas in spiral galaxies show a shift 
$z_{\rm tot}$ in their intrinsic spectra which includes the contributions of:
1) the cosmological expansion
(recession of galaxies), 2) the peculiar motion of the galaxy,
3) the thermal motion of atoms within the stars and gas, 4) the
gravitational field within the galaxy and stars, and finally 5)
the motion of the stars around the galactic center.

When the ``contaminating'' effects from 1)-3) are subtracted
from the data, usually astronomers report the resulting $z$ in terms of a
velocity field $v$. Nevertheless it is instructive to make the analysis in
terms of the  quantities that are
most directly observable: the $z$'s. We perform this in order to keep track of the effect of
the underlying assumptions,
and to enable us to carry the analysis  when these are not longer valid, as
will be the case in
some models we  will consider.  \\
\\
The starting point is to assume that stars behave like test particles
which follow geodesics  of  a static and spherically
symmetric space-time associated with sources that we do not
specify for the moment.  The most general  line element
of the space-time  in these circumstances takes  the form:
\be
ds^2 = -N^2(r) dt^2 + A^2(r)dr^2 + r^2d\theta^2
+ r^2\sin^2\theta d\varphi^2 \,\,\,\,\,.
\label{RGMS}
\ee
Next we consider two observers ${\cal O}_E$ and ${\cal O}_D$ with
four velocities $u^\mu_E$, $u^\mu_D$ respectively. Observer
${\cal O}_E$ corresponds to the light emitter (i.e., to the stars
placed at a point $P_E$ of space-time), and ${\cal O}_D$ represents
the detector at point $P_D$ located far from the emitter and that
can be idealized to  correspond to  ``spatial infinity''.

Without loss of generality, we can assume that the stars move on the
galactic plane $\theta=\pi/2$, so that
$u^\mu_E = (\dot{t}, \dot{r}, 0, \dot{\varphi})_E$,
where the dot stands for derivation with respect to the proper time of the
particle.  \\
\\
On the other hand, we suppose that the detector is
static (i.e, ${\cal O}_D$'s 4-velocity
is tangent to the static Killing field ${{\partial} \over {\partial t}}$),
and so
with respect to the above coordinates its 4-velocity is
$u^\mu_D= (\dot{t}, 0, 0, 0)_D$.

As usual, the consideration of the norm of the four velocity $(u^\mu u_\mu=-1)$, gives,
\\
\be\label{mg}
-1 = - N^2(r) (\dot{t}^2) +
A^2 (\dot{r}^2) + r^2 (\dot{\varphi}^2) \,\, ,
\ee

The energy and the angular momentum per unit of mass
at rest of the test particle
are conserved quantities and can be written as
\\
\be\label{cc}
E = - g_{\mu\nu} \varepsilon^{\mu} u^{\nu} =
N^2 (\dot{t}) \,\, , \quad
L = g_{\mu\nu} \psi^{\mu} u^{\nu} = r^2 (\dot{\varphi})\,\, ,
\ee
where $\varepsilon^{\mu}$, $\psi^{\mu}$ denote the
timelike and rotational killing fields of the metric
(\ref{RGMS}) respectively. Introducing these constants of motion
in the line element (\ref{mg}), we obtain
\be\label{egrpunto}
{N^2}{A^2}
(\dot{r})^2 + N^2 \left[ \frac{L^2}{r^2} + 1 \right]
= E^2  \,\, ,
\ee

This equation
shows that the radial motion of a geodesic is the same as that
of a particle with position dependent mass and with energy $E^2/2$
in ordinary non-relativistic mechanics moving in the effective potential
\be
V_{eff} (r) = N^2(r)
\left[ \frac{L^2}{r^2} + 1 \right].
\label{veffe}
\ee
\\
As we mentioned, the RC of spiral galaxies
are inferred from the red and blue shifts of the
emitted radiation by stars moving in ``circular orbits''
on both sides of the central region \cite{rubin1}.
The light signal travels on null geodesics with tangent
$k^{\mu}$. We may restrict
$k^{\mu}$ to lie also in the ``equatorial plane'' $\theta = \pi/2$,
and evaluate the frequency shift for a light
signal emitted from ${\cal O}_E$ in circular orbit and
detected by ${\cal O}_D$. The conditions for circular orbits
 $\partial_r V_{eff}=0$ and $\dot r=0$ lead to
\bea\label{L}
L^2 &=& \frac{r^3\partial_r N/N}{1- r\partial_rN/N}\,\,\,,\\
\label{En}
E^2 &=& \frac{N^2}{1-r\partial_r N/N}\,\,\,\,.
\eea
\\
The frequency shift associated to the emission and detection is given by,
\be\label{z}
z= 1- \frac{\omega_E}{\omega_D} \,\,\,\,,
\ee
where
\be\label{freq}
\omega_C=  -k_\mu u^\mu_C|_{P_C}\,\,\,\,,
\ee
the index $C$ refers to emission or detection at the corresponding 
space-time point. \\
\\
Two frequency shifts corresponding to a
maximum and minimum values  are associated with light
propagation in the same and opposite direction of motion of the emitter
respectively (i.e., $k^r=k^\theta=0$). Such shifts are 
the frequency shifts of a receding or
approaching star respectively. Using
 the constancy along the geodesic of the product
of the Killing field ${{\partial} \over {\partial t}}$ with a geodesic tangent
together with (\ref{freq}) and (\ref{z}), and expressions (\ref{En}) and 
(\ref{L}), we find the two shifts to be,
\be
z_{\pm} =
1 - \frac{N_D}{N(r)}\frac{\left( 1 \mp
\left[ r \partial_r N(r)/N(r) \right]^{1/2} \right)}
{\left[1 - r\partial_r N(r)/N(r)\right]^{1/2}}
\,\,,
\label{redshift1}
\ee
where $N(r)$ represents the value of the
metric potential at the radius of emission $r$, and $N_D$ the
corresponding value of $N(r)$ at $r\rightarrow \infty$ where
the detector is supposed to lie.
For asymptotically flat space-times $N_D\rightarrow 1$.
However, for space-times
generated by global-monopoles $N_D\rightarrow (1-\alpha)^{1/2}$ (see the
section VI).  \\
\\
It is worth noting that in terms of the tetrads 
$e_{(0)} = N^{-1}{\partial\over {\partial t}},
e_{(1)} = A^{-1}{\partial \over {\partial r}},e_{(2)} = r^{-1}{\partial \over
{\partial \theta}},e_{(3)} = (r\sin \theta )^{-1}{\partial\over {\partial \phi}}$,
the frequency shifts take the form,
\be\label{ztf}
z_\pm= 1 - \frac{N_D}{N}\left(1\mp v \right) \Gamma\,\,\,,
\ee
where $v:= (\sum_{i= 1,2,3}(u_{(i)}/u_{(0)})^2)^{1/2}$ and $u_{(\mu)}$ 
stands for the components of the star's four
velocity  along the tetrad (i.e., the velocity
measured by an Eulerian observer whose world line is tangent to the  static
Killing field) and  $\Gamma= (1-v^2)^{-1/2}$ is the usual  Lorentz factor. 
Clearly, in the present case of circular orbits on the plane 
$\theta=\pi/2$, it results that $v=u_{(3)}/u_{(0)}\equiv r\partial_r N(r)/N(r)$.
\\
It is convenient to define the quantities:
  $z_D = \frac{1}{2} (z_+ - z_-)$ and$z_A = \frac{1}{2} (z_+ + z_-)$  which
are easily connected to the
observations. From the expression (\ref{redshift1}) we obtain,
\be
z_{D} (r) = \frac{N_D}{N(r)}\frac{\left[ r \partial_r N(r)/N(r)\right]^{1/2}}
{\left[ 1- r \partial_r N(r)/N(r)\right]^{1/2}}   \,\,.
\label{rsm}
\ee

\be
z_{A} (r) =  1- \frac{N_D}{N(r)}\frac{1}
{\left[ 1- r \partial_r N(r)/N(r)\right]^{1/2}}   \,\,.
\label{rsm2}
\ee
\\
We note, for example, that $ (z_A-1)^2- z_D^2 = (N(r)/N_D)^{-2}$, 
and thus we could in principle recover $N(r)$ directly from the
observations. Then we can use this $N(r)$ to recalculate 
$z_A$ and $z_D$ from  the
above expressions and compare them  with the
measured values. This would be a test of the assumption that the dynamics is
determined by the geodesics of a
stationary metric, quite independently of the assumption of the dynamics of
the geometry itself or for of the nature
of the dark matter.

\section{Gravitational field in the dark matter zone}

In this section we will use the form of the RC to
obtain the spacetime metric and information on
 the matter content. The energy-momentum tensor must be diagonal and
spherically symmetric, as dictated by the
symmetries of the space-time (\ref{RGMS}), so we define:
\be
\rho \equiv - T^t_t    \,\,\,\,\,,\,\,\,\,\,\,\,\,
P_r \equiv T^r_r   \,\,\,\,\,,\,\,\,\,\,\,\,\,
P_{\theta} \equiv T^{\theta}_{\theta} = T^{\varphi}_{\varphi}  \,\,\,\,\,.
\label{preden}
\ee
and $T^\mu_\nu=0$ for $\mu\ne\nu$.

We will for convenience introduce the following alternative form
 of the metric variables:
\be
A^{2}(r) =
\left(1 -
\frac{2 m(r)}{r}\right)^{-1}
\,\,\,\,\,, \,\,\,\,\,
N^2(r) = \exp[2 \nu(r)]   \,\,\,\,\,,
\label{AADM}
\ee
Einstein's equations  then read
\bea
\label{massudm}
\partial_{r} m &=&  4\pi r^2 \rho \,\,\,\,\,,\\
\label{lapsefidm}
\partial_{r} \nu  &=& \frac{m(r)}{r^2}
\left(1 - \frac{2m(r)}{r}\right)^{-1}
\left(1 + \frac{4\pi r^3 P_{r}}{m(r)}\right) \,\,\,\,,
\eea
The equation of hydrostatic equilibrium resulting from the
conservation of the energy-momentum tensor $\nabla_{\nu} T^{\mu \nu} = 0$
becomes
\be
\label{fluidofi}
\partial_{r} P_{r} =
- (\partial_{r} \nu) \rho
\left(1 + \frac{P_{r}}{\rho}\right)
- \frac{2}{r \rho} (P_{r} - P_{\theta}) \,\,\,\,\,.
\ee

We note that  the observations in spiral galaxies \cite{rubin1}, 
yield $z_D =v \approx constant$ and $z_D>> z_A$.
From these conditions and from (\ref{rsm}) we obtain
\be
 \frac{1}{N(r)}\frac{1}
{\left[ 1- r \partial_r N(r)/N(r)\right]^{1/2}}  \approx 1 \,\,.
\label{redshiftflat}
\ee

and
\be
v \equiv \left[ r\partial_r N(r)/N \right]^{1/2}  \,,
\label{redshiftflatone}
\ee
the value of $v$ roughly ranges from
$10^{-4}$ to $10^{-3}$ depending of a particular spiral galaxy.
The integration of (\ref{redshiftflat}) gives,
\be
N(r) = \left(\frac{r}{r_g}\right)^{v^2}\,\,\,\,,
\label{gtt}
\ee
where $r_g$ is constant. \\
\\
Note that using (\ref{gtt}) in
Eqs. (\ref{lapsefidm}) and (\ref{fluidofi}), we obtain a system of
three equations [i.e., Eqs.(\ref{massudm})$-$(\ref{fluidofi})] 
for four unknowns (i.e, $m,\rho,P_r,P_\theta$). In the case of a 
perfect fluid, however, the four unknowns are reduced to three 
since $P_r=P_\theta\equiv p$. This therefore constrains the equation of state 
$p=p(\rho)$. On the other hand, 
for the case where the matter content is associated for example, with
a scalar field, then $\rho$, $P_r$ and $P_\theta$ are not independent but 
are given in terms of the gradients of the field. This constrains the 
form of the scalar potential.

We will look for a solution that satisfies the Newtonian conditions,
\\
\begin{eqnarray}
\label{cond2}
P_{r, \theta} << \rho  \,\,\,\,\,,\\
\label{cond3}
m(r) << \frac{r}{2}   \,\,\,\,\,,\\
\label{cond4}
4\pi r^3 P_{r} << m(r)   \,\,\,\,\,.
\end{eqnarray}
\\
Under these conditions the Eqs.
(\ref{massudm})$-$(\ref{fluidofi}) reduce to 
\\
\begin{eqnarray}
\label{massu1}
\partial_{r} m(r) &=&  4\pi r^2 \rho \,\,\,\,\,,\\
\label{lapsefi1}
\partial_{r} \nu  &=& \frac{m(r)}{r^2}  \,\,\,\,, \\
\label{fluidofi1}
\partial_{r} P_{r} &=&
- (\partial_{r} \nu) \rho   \,\,\,\,\,,
\end{eqnarray}
\\
using the expression (\ref{gtt}) for $N$ in (\ref{lapsefi1}),
(\ref{fluidofi1}) with $\nu={\rm ln[N]}$, and solving the system, we obtain
\\
\begin{eqnarray}
m(r) \approx  v^2 r,  \,\,\,\,\,
\rho (r) \approx \frac{v^2}{4\pi r^2},  \,\,\,\,\,
P_r \approx \frac{v^4}{8\pi r^2},  \,\,\,\,\,.
\label{solapro3}
\end{eqnarray}
\\
The solution corresponds to the relation $P_r\approx v^2 \rho/2$ which
looks somewhat peculiar.
If we view this as the equation of state of a perfect fluid, in 
the case of an ideal gas we would
conclude that its temperature $T$ is constant and proportional to $v^2$.
 The interpretation is that the dark matter represented by a perfect fluid 
is made of particles (ideal gas) that interact 
among themselves strongly enough to maintain thermal equilibrium but do
not interact in the same way with ordinary matter or with photons.
The idea is then that the dark matter temperature determines its density
profile and 
the spacetime metric, and through this, the rotation curves of the stars in
the galaxy. One of the problems of this type of models is the need to 
explain why the isothermal configuration of the fluid does not extend 
to the center of the galaxy.
\\
Using Eqs. (\ref{gtt}) and (\ref{solapro3}) in (\ref {RGMS}) we obtain
the final result for the metric in this limit:
\\
\begin{equation}
ds^2 = - \left(\frac{r}{r_g}\right)^{2v^2} dt^2 + \left(1 -
\alpha \right)^{-1} dr^2 + r^2 (d\theta^2
+ \sin^2\theta d\varphi^2) \,\,\,\,\,,
\label{metrica}
\end{equation}
\\
where $\alpha = 2 v^2$. We emphasize  that this solution is only valid
in the flat RC zone.  We know that eventually this must be
matched to a 
Robertson-Walker metric describing the
universe or alternative we might use the asymptotically flat idealization
for regions very far from the galaxy
in question. 
To do so we must consider (\ref{metrica}) as describing the spacetime
geometry for
$r<R_0$ and the Schwarzschild metric for
$r>R_0$, where
$R_0$ is the radius where  the flat rotation curves end.
The advantage of this approximation is that far from the galaxy
the spacetime is Minkowskian  a fact that facilitates, for example, the
analysis of the propagation of light signals.
\\
Matching continuously the two metrics at $R_0$ allows the determination
of the integration constants:
\\
\bea
ds^2 = - \left(1 - \alpha \right)
\left(\frac{r}{R_0}\right)^{2v^2} dt^2 + \left(1 -
\alpha \right)^{-1} dr^2 + r^2 (d\theta^2
+ \sin^2\theta d\varphi^2)   \,\,\,\,\,\,\,\,\,  r < R_0  \nonumber \\
ds^2 = - \left(1 - \frac{2 M}{r}\right) dt^2 +
\left(1 - \frac{2 M}{r}\right)^{-1} dr^2 + r^2 (d\theta^2
+ \sin^2\theta d\varphi^2)   \,\,\,\,\,\,\,\,\,  r > R_0
\label{metricaone}
\eea
\\
where $M = \alpha R_0/2$.

Here we are taking the view that the region of flat rotation curves,
terminates in a narrow transition region
where the behavior of the density changes from the $1/r^2$ to a constant
that for simplicity we take to be zero.
The point being that in the limit in which  the region is very narrow the
metric will not change abruptly as
we cross the region, but the derivative of the metric coefficients will
experience a jump. It would be
interesting to consider  various  alternatives for the details of  the
interpolating regions. \\
\\
An alternative to the search of solutions satisfying the Newtonian 
conditions (\ref{cond2})$-$(\ref{cond4}) is to 
start from the Eqs.. (\ref{massudm})$-$(\ref{fluidofi}) without 
imposing the Newtonian approximation but assuming that the dark matter is 
represented by a perfect fluid. Then again, the use of (\ref{gtt}) with 
$P_r=P_\theta$ allows the integration of (\ref{massudm})$-$(\ref{fluidofi}) 
as follows:
\bea
m(r) = \frac{v^2}{2}
\frac{(2 - v^2)}{(1 + 2v^2 - v^4)} r  \,\,,\,\,\,
\label{sol1}
\rho (r) = \frac{v^2}{8\pi}
\frac{(2 - v^2)}{(1 + 2v^2 - v^4) r^2}  \,\,,\,\,\,
\label{sol2}
P_r = \frac{v^4}{8\pi}
\frac{1}{(1 + 2v^2 - v^4) r^2}  \,\,\,\,\,,
\label{sol3}
\eea
which results in an ``equation of state'' $P_r= v^2 \rho/(4-2v^2)$.
In practice $v^2 << 1$, so we recover the solution (\ref{solapro3}).

At this point we can check whether the approximations we considered are
self consistent.
That is, we substitute (\ref{gtt}) in the left hand side of
(\ref{redshiftflat}) obtaining:
\be
 \frac{1}{N(r)}\frac{1}
{\left[ 1- r \partial_r N(r)/N(r)\right]^{1/2}}=\sqrt{{1}\over
{(1-v^2)(1-\alpha)}} (r/R_0)^{-v^2} \,\,.
\label{rsm3}
\ee
 The difference between this expression and $1$ must be negligible  in
comparison to $v$ which itself is of
the order $10^{-3} - 10^{-4}$. This requires $r/R_0$ to be neither to
large nor to small. To get an estimate
we use the approximation $ X^{-v^2} \approx 1 - v^2 {\rm ln}(X)$ valid for 
$ |v^2 {\rm ln}(X)| <<1$.
These requirements are then:
\be
| {\rm ln} (r/R_0)| << v^{-2}
\label{}
\ee
Thus the approximations are self consistent as long as $- 10^{6} <<{\rm ln}
(r/R_0)<<10^{6}$, which
doesn't impose any practically relevant constraints 
for the case of the galaxies and
the extent of the flat RC.  \\
It is worth noting that the form of the metric that we have obtained
differs from what would be naively
expected:
\\
\begin{equation}
ds^2 = - \left(1 + 2
\Phi \right) dt^2 + \left(1 +
2\Phi \right)^{-1} dr^2 + r^2 (d\theta^2
+ \sin^2\theta d\varphi^2) \,\,\,\,\,,
\label{metricanewto}
\end{equation}
\\
with $\Phi $ representing the Newtonian potential. This form
is often implicitly assumed (see\cite{peebles}) and the fact that it is
not appropriate for the
region where the RC are flat lies at the core of the problems with the
analysis of \cite{Edery} (see \cite{Bek}).  \\
\\
We also point out that if we assume that the flat RC extend indefinitely, 
the resulting spacetime is not
asymptotically flat but rather is asymptotically flat but for a deficit angle
(AFDA) \cite{ulises}.
In this context we stress that it is possible to consider
such ``idealized infinitely extended
galaxies'' as isolated objects in the framework of general relativity 
by replacing the asymptotically-flat framework by the
framework AFDA \cite{ulises}. One might want to embark in such considerations
since in fact the RC remain flat to the
farthest distances that can be observed. On the other hand there is a
natural way to estimate an upper
bound for the cutoff of such behavior. The idea is to  consider the point
at which the decaying density
profile associated with the galaxy becomes smaller than the average energy
density of the universe. We call this bound 
$R_{Max}^{U}$. The value for $R_{Max}^{U}$ is obtained 
by imposing the condition that the density at this point, provided by 
Eq. (\ref{solapro3}), is to coincide with the mean density 
of the universe: 
\begin{eqnarray}
\rho (R_{Max}^{U}) \approx \frac{v^2}{4\pi (R_{Max}^{U})^2} = 
\rho_{U}  
\,\,\,\,\,,
\label{estimate1}
\end{eqnarray}
\\
where $\rho_{U}$ is the mean density of the universe. Then we have
\\
\begin{eqnarray}
R_{Max}^{U} = \sqrt{\frac{v^2}{4\pi \rho_{U,I}}} \,\,\,\,\,.
\label{estimate2}
\end{eqnarray}
\\
Now we introduce the value $\rho_{U}$ in terms of the dimensionless 
Hubble parameter $h$ defined as $H_0 \equiv 100 h \,{\rm km/(sec\,Mpc)}$ and 
$\Omega_{U} \equiv \rho_{U}/\rho_{\rm crit}$
\\
\begin{eqnarray}
\rho_{U} = 1.8791 \times 10^{-26} \Omega_{U} h^2 \, {\rm kg/m^3} = 
2.78 \times 10^{11} \Omega_{U} h^2 M_{\odot}/{\rm Mpc^3} 
\,\,\,\,\,.
\label{estimate3}
\end{eqnarray}
\\
We obtain, 
\\
\begin{eqnarray}
R_{Max}^{U} = 2.45 \times 10^6 \,\frac{v}{h}\, \Omega_{U}^{-1/2}\, {\rm Kpc}
\,\,\,\,\,.
\label{estimate4}
\end{eqnarray}
\\
Taking $h = 0.65$ and 
$v \approx (10^{-3}-10^{-4})$, we have
\\
\begin{eqnarray}
R_{Max}^{U} = 3.77 \times (10^{3}-10^{2}) \Omega_{U}^{-1/2}\, {\rm Kpc}\,\,\,\,\,.
\label{estimate5}
\end{eqnarray}
Moreover, for $\Omega_U = 1$,
\\
\begin{eqnarray}
R_{Max}^{U} \approx 3.77 \times (10^3-10^2)\, {\rm Kpc} \,\,\,\,\,.
\label{estimate9}
\end{eqnarray}
\\
On the other hand, the measured flat regions are about 
$R_0 \approx 2 R_{\rm opt}$ where $R_{\rm opt}$ is the radius encompassing
83 per cent of the total integrated light of the galaxy. 
We can take as a typical value $R_{0} \approx 30 {\rm Kpc}$, 
then $R_0 < R_{Max}^{U}$.

\section{Bending of light in the Dark Matter zone}
\bigskip

One of the ways we could in principle explore 
the issue of whether the flat RC are the
result of some
form of unknown matter or the result of the change in the 
dynamical laws that governs the motion of particles, 
would be by studying the light deflection by galaxies. In particular, 
by studying the deflection of photons passing through the  region where the
RC are flat. Let us thus consider a photon
approaching the spiral galaxy from far distances. We will compute the
bending of light assuming the metric
that has been matched with an asymptotically flat exterior, i.e., Eq. 
(\ref{metricaone}).

\begin{figure*}
\vspace{1cm}
\psfig{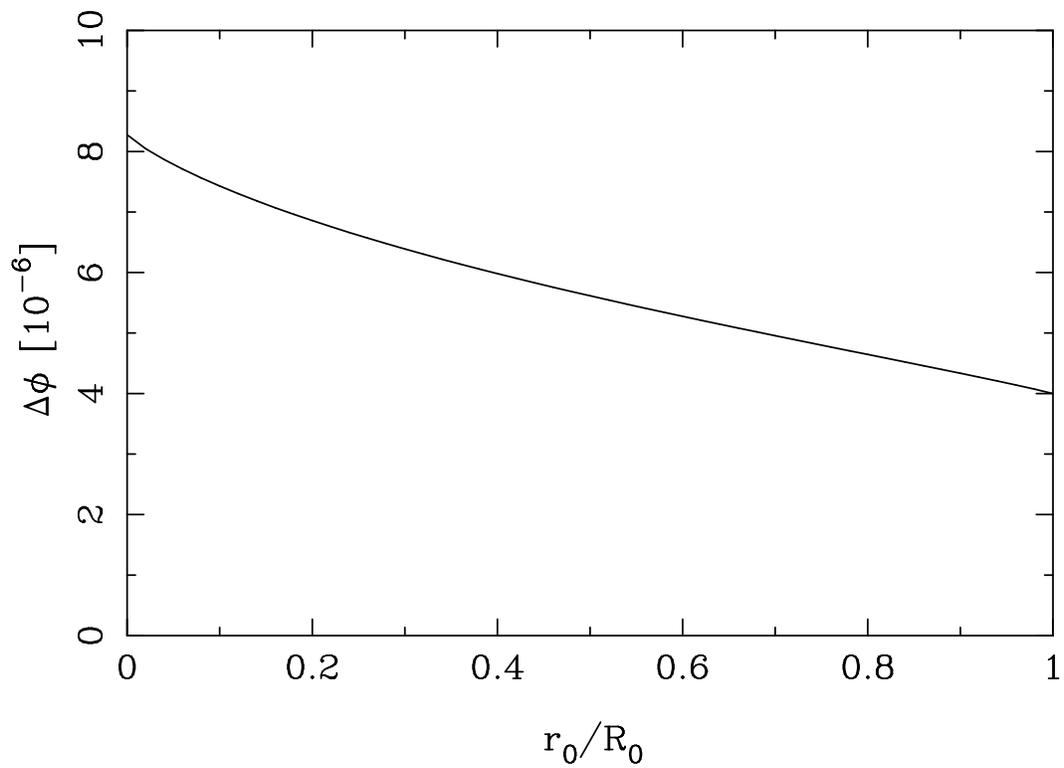}
\hspace*{-2.5in} %%\vskip 2cm
\caption[]{\label{f:bending2}
Bending angle of the light as a function of parameter $r_0/R_0$.}
\end{figure*}

The bending of the light results \cite{weinberg},
\\
\be
\Delta \phi = 2 |\phi(r_0) - \phi_{\infty}| - \pi
\label{resul}
\ee
\\
where $\phi_{\infty}$ is the incident direction and $r_0$ is the coordinate 
radius of closest approach to the center of the galaxy.
\\
\be
\phi(r_0) - \phi_{\infty} = \int_{r_0}^{\infty} A(r)
\left[ \left( \frac{r}{r_0} \right)^2 \frac{N^2(r_0)}{N^2(r)} - 1
\right]^{-1/2}
\frac{dr}{r}  \,\,\,\,\,\,\,\,.
\label{resul0}
\ee
\\
The integral is split in two parts for the two domains of metric
(\ref{metricaone}):
\\
\be
\phi(r_0) - \phi_{\infty} = \int_{r_0}^{R_0} A(r)
\left[ \left( \frac{r}{r_0} \right)^2 \frac{N^2(r_0)}{N^2(r)} - 1
\right]^{-1/2}
\frac{dr}{r} +
\int_{R_0}^{\infty} A(r)
\left[ \left( \frac{r}{r_0} \right)^2 \frac{N^2(r_0)}{N^2(r)} - 1
\right]^{-1/2}
\frac{dr}{r}    \,\,\,\,\,\,\,\,.
\label{integralone}
\ee
\\
The second integral is computed by expanding the integrand in powers of
$M/r_0$ and $M/r$ \cite{weinberg} using (\ref{metricaone}) for $r>R_0$,
\\
\be
\left[ \left( \frac{r}{r_0} \right)^2 \frac{N^2(r_0)}{N^2(r)} - 1 \right] =
\left( \frac{r}{r_0} \right)^2 \left[ 1 +
2M \left( \frac{1}{r} - \frac{1}{r_0} \right) + ....\right]  \nonumber \\
= \left[ \left( \frac{r}{r_0} \right)^2 - 1 \right]
\left[1 - \frac{2Mr}{r_0 (r + r_0)} + ....\right] \,\,\,\,\,\,,
\ee
\\
and the results is
\bea
&& \int_{R_0}^{\infty} A(r)
\left[ \left( \frac{r}{r_0} \right)^2 \frac{N^2(r_0)}{N^2(r)} - 1
\right]^{-1/2}
\frac{dr}{r}
= \int_{R_0}^{\infty}
\frac{dr}{\left[ \left( \frac{r}{r_0} \right)^2 - 1 \right]^{1/2}}
\left[ 1 + \frac{M}{r} + \frac{Mr}{r_0 (r + r_0)} + ....\right] \nonumber \\
 &=& \arcsin \left(\frac{r_0}{R_0}\right) + \frac{M}{r_0} \left[
2 - \left[ 1 - \left(\frac{r_0}{R_0}\right)^2 \right]^{1/2} -
\left( \frac{R_0 - r_0}{R_0 + r_0} \right)^{1/2} \right] + .... \,\,\,\,\,\,.
\label{resul1}
\eea
\\
The first integral of (\ref{integralone}) with (\ref{metricaone})
for $r<R_0$ gives
\\
\bea
\int A(r)
\left[ \left( \frac{r}{r_0} \right)^2 \frac{N^2(r_0)}{N^2(r)} - 1
\right]^{-1/2}
\frac{dr}{r} &=& (1 - \alpha)^{-1/2}
\int \left[ \left( \frac{r}{r_0} \right)^{2(1-v^2)} - 1 \right]^{-1/2}
\frac{dr}{r} \nonumber \\
&=& \frac{(1 - \alpha)^{-1/2}}{(v^2 - 1)} \arctan
\left[ \left( \frac{r}{r_0} \right)^{2(1-v^2)} - 1 \right]^{-1/2}
\,\,\,\,\,\,.
\label{resul2}
\eea
\\
Finally, using (\ref{resul1}) and (\ref{resul2}) in (\ref{resul}) and then 
using (\ref{resul0}), the bending angle of light yields,
\\
\bea
\Delta \phi
&=& \left| 2\arcsin \left( \frac{r_0}{R_0} \right) + \frac{2M}{r_0}
\left[ 2 - \left[ 1 - \left(\frac{r_0}{R_0}\right)^2 \right]^{1/2} -
\left( \frac{R_0 - r_0}{R_0 + r_0} \right)^{1/2} \right] \right. +
\nonumber \\
&& \left. \frac{2 (1 - \alpha)^{-1/2}}{(v^2 - 1)} \left(  \arctan
\left[ \left( \frac{R_0}{r_0} \right)^{2(1-v^2)} - 1 \right]^{-1/2}
- \frac{\pi}{2} \right)\right| - \pi
\,\,\,\,\,\,,
\label{resul3}
\eea
\\
where we took the limit
\\
\be
{\rm lim}_{r \rightarrow r_0} \arctan
\left[ \left( \frac{r}{r_0} \right)^{2(1-v^2)} - 1 \right]^{-1/2} =
\frac{\pi}{2}
\,\,\,\,\,\,.
\label{resul4}
\ee
\\
If we put $r_0 = R_0$ in (\ref{resul3}), we obtain the standard 
result for the 
Schwarzschild metric with mass $M = \alpha R_0/2$ and with 
$\Delta \phi = (4M/r_0) = 4 \times 10^{-6}$.
Fig.\ref{f:bending2} shows the bending angle of light 
$\Delta \phi$ as a function of the parameter $r_0/R_0$.
If we take the impact parameter, $r_0$, to be in the range
of the measured flat regions by neutral Hydrogen measurements (HI): 
$R_{\rm opt} \leq r_0 \leq 2R_{\rm opt} = R_0$, then we have 
$1/2 \leq r_0/R_0 \leq 1$. In this case, the maximum value for 
$\Delta \phi$ is obtained for the value $r_0/R_0 = 1/2$. Recently,
the investigations for determining the radius of dark matter halos
have gone beyond the HI measurements using satellite galaxies \cite{smith}
or the weak lensing of background galaxies by foreground dark halos 
\cite{smail}. From these measurements, halo radii of more than 200 {\rm Kpc}
are inferred. For our galaxy 230 {\rm Kpc} \cite{kulessa} and from 
satellite galaxies of a set of spiral galaxies show more than 
400 {\rm Kpc} \cite{frenk}. By taking 
$R_0 \approx 230$ {\rm Kpc}, we would have 
$R_{\rm opt} \leq r_0 \leq 15 R_{\rm opt} = R_0$
(where we have chosen $R_{\rm opt} = 15 {\rm Kpc}$). 
In this case we would have $1/15 \leq r_0/R_0 \leq 1$ and 
a value near to the maximum in the fig.\ref{f:bending2}. 
It would be interesting to explore the possibility to have 
relevant observations in this context.\\
\\
We discuss next two of the three simplest scenarios in which the dark
matter corresponds to coherent scalar fields.

\section{Spontaneous scalarization}

As we mention in the introduction,
the phenomenon of spontaneous scalarization in compact objects (notably
in neutron stars) that arise in
a class of scalar tensor theories of gravity \cite{Damour} is the other
mechanism that allows the appearance of a non trivial scalar field in the
absence of a direct coupling between the scalar field  and ordinary matter.
The general feature of these kind of theories is a scalar field coupled
non-minimally to gravity which leads to an effective
gravitational coupling which depends explicitly on the scalar field.
The non-trivial scalar field configuration 
appears when the object is compact enough so that the
energy of the configuration for a fixed baryon number is minimized 
through a change in the value of the effective gravitational constant. 
That is, for a fixed baryon number, the energy of 
the configuration with a scalar field is lower than the corresponding 
configuration in absence of a scalar field \cite{SSN}. An heuristic
interpretation that is confirmed by the numerical results shows that, from a
Newtonian point of view, the relevant quantity to be minimized is the
combination $GM$ instead of the total mass $M$. We observed that although
the obvious additional contributions to $GM$ \cite{footnote} are both 
positive and thus increasing the value of $GM$,
their effect is more than compensated by the reduction of the value of 
the contribution $GM_{\rm bar}$, which is the leading term in $GM$. 
Thus there appears a nontrivial configuration of the scalar field 
which is associated with the minimization
(at fixed total baryon number) of the
the value $GM$ \cite{SSN}.  \\
\\
Several problems arise if we want to use this mechanism to induce a 
non trivial configuration of
a scalar field at the galactic scale. 
First, in the model studied so far we have seen that
spontaneous scalarization occurs only if
the object is compact enough, that is, if $GM/R \sim 1/2$ and needless to
say that
the galaxy as a whole does not satisfy this criteria 
(except perhaps at the center). If we assume that a large dense
object lies at the center of the
galaxy  one would need some very unusual equation of state to
overcome the standard limits on  the mass of these objects associated with
the requirement of stability against
collapse.  But even if we were to assume such an object, the scalar field
associated with the phenomena of spontaneous scalarization falls as 
$1/r$ (at least in the models considered so far) so it would not be
relevant at the distances associated with the flat RC that lie at a
distance of the order of kiloparsecs from
the galactic center. Finally, the energy of the configurations with
nontrivial configuration according to this phenomenon is 
smaller than that of the corresponding configuration in which 
the scalar field vanishes,
thus the phenomenon seems to take us  in the  opposite direction  as
compared to what seems to be
required to
explain the additional attractive effect on the test stars in the galaxy. 
If we wanted to consider extended
objects other that neutron stars, it is not even clear how to build a
sufficiently dense object. The only
possibility would seem to be boson stars \cite{MFS} which now would act only 
as triggers
of the spontaneous scalarization.
These models would require to hypothesize two scalar fields, one providing
the oscillating boson field
of the boson star, and a second one providing the mechanism for spontaneous
scalarization.
On the other hand we must point out that although boson star masses are
usually very small, when one
introduces self-interactions
their mass can be as large as $10^{27}
\lambda^{1/2} M_\odot$ (for a scalar field mass 
$m\sim 10^{-5}$ eV and a sufficiently 
large self-interaction constant $\lambda$)\cite{Colpi}. Still we face 
the problem associated with the rapid fall off of the energy
density associated with the scalar field, which would go like $1/r^4$. \\
\\
The hope here would be to consider alternative forms of the nonminimal
coupling, with the possible introduction
of  various forms of self interaction terms for the scalar field, that
would not only lead to spontaneous
scalarization but to a rather different fall off behavior of the scalar field. 
Nevertheless, as was already mentioned, there is one very serious problem
remaining with this type of
scenario, and it is the issue of black holes.
 There is at the present time mounting evidence that there is at the center
of most galaxies a
very massive black
hole, and in view of the no hair theorems for scalar fields 
\cite{MH,DS,JDB}, it
seems clear that the phenomena
of spontaneous scalarization does not have an analogy when the compact
objects are replaced by black holes.
Thus in those galaxies the scalar field would relax to the trivial
configuration and thus any explanation of
the RC based on that phenomena would cease to be operative. There are, 
however, some small loopholes
remaining in the black hole uniqueness theorems for the case of nonminimally
coupled scalar fields which leave a ray of hope in this 
general direction, and which are currently under investigation\cite{Igor2}.  \\

For the case of massless bosons (massless complex scalar fields),
a Newtonian analysis leads to
flat RC \cite{schunk}. Unfortunately, in that work, 
the author neglect to note that the 
RC are not directly observable but only inferred from the corresponding 
light shifts. As it turns out, in that model 
the ``gravitational'' red-shift 
would be very large to the point that by ignoring it, 
the author is ignoring effects of the 
same order of magnitude as the ones that are being considered. 
Moreover, the law of composition of velocities used there to 
reproduce the RC is not valid.

In the following section we analyze the case for the
matter represented by global monopoles non-minimally coupled to
gravity.

\section{Global Monopoles}

We will now consider one example of what we feel is at this time the most
promising
class of models: nonminimally coupled global monopoles. The main
results of this section have been reported in
\cite{NSS}.  \\
\\
Particle physics models predict the formation of
topological defects during phase transitions in the
early universe. The mechanism argued for the formation
of these is the spontaneous breaking of symmetry
of the model under consideration leading to a manifold of degenerate
vacua with nontrivial topology. Topological defects
can be classified according to the topology of the
vacuum manifold. If the manifold of equivalent vacua,
${\cal M}$, contains unshrinkable surfaces, $\pi_{2}({\cal M})\neq I$,
then monopoles are formed. These can be classified into
local and global monopoles depending on whether the
symmetry broken is local or global. In the first case
(gauge monopoles) the monopole configuration has finite
energy concentrated in a small core and produces an
asymptotically flat spacetime, while in the second case, 
the global-monopole configuration 
has a linearly divergent energy due to the long range
Nambu-Goldstone field with energy density decreasing
with the distance as $r^{-2}$. As we have mentioned,
this behavior is very appealing in view of the fact that
this is precisely what seems to be required in a naive
picture to provide a natural explanation for the flatness of
the RC.  \\
\\
It was shown by
Barriola and Vilenkin\cite{vilenkin1} that this
linearly divergent ``mass'' has, at large distances, 
an effect analogous to that of a deficit solid angle $\alpha$ plus
that of a tiny mass associated to the core of the monopole. Then, assuming the
existence of a global monopole
in a typical galaxy the total Newtonian mass contribution
of the portion of the global monopole contained within $r_{\rm gal}$ (with
$r_{\rm gal}\approx 15$ Kpc) is found to be
$M\sim \alpha r_{\rm gal}/2 \approx 10^{69}$ GeV, where we took
a typical grand unified value $\eta\approx 10^{16}$ GeV, 
and where $\alpha= 8\pi G\eta^2$. This estimate turns out to
be 10 times the total mass due to the contribution of $10^{11}$ solar-mass
in a typical galaxy (i.e., $M_{\rm stars}\approx 10^{68}$ GeV). These numbers
are again what is needed to account for the observations. Finally, if we
assume that the field of the monopole extends on average a distance of ten
galactic radii from the galaxy where the configuration presumably coincides
with that of the monopole centered in the neighbouring galaxy, then
$M\approx 10^{70}$ GeV, which is 100 times that of the galaxy. This value
leads to a contribution of the monopole to the total average density in the 
universe, which is of the order of magnitude
predicted by the standard inflationary scenarios. Actually, it is
the reversed argument that helps to place upper bounds on the density number
of global monopoles present in the universe \cite{Hiscock}. On the 
other hand, Harari and Loust\'o \cite{harari}, showed that the small
effective mass $m_{\rm core}\approx 0.8\alpha $ is in fact
negative and produces a repulsive potential.
They studied the motion of test particles in the spacetime of a
global monopole concluding that there are no bound orbits.
This result led thus to the unavoidable conclusion that minimally
coupled global monopoles are not good candidates to explain the RC despite the
suggestive numbers and features considered above.  \\
\\
Another problem is the fact that
the monopole configuration is rather unique,
in the sense that it is basically independent of
the ordinary matter content in the corresponding galaxy,
which conflicts with the fact that there is
a rather large range of galactic masses for which the dark matter
component is about ten times more massive than the ordinary matter
component \cite{wald}. \\
\\
Recently, we have shown \cite{NSS}, 
that coupling global monopoles non-minimally
to gravity is possible to avoid the most unwanted features
of the minimal case, notably, the lack of
bound orbits, and the universality of the monopole configuration. \\
\\
Specifically we considered a theory of a triplet of scalar fields
$\phi^a$, $a=1,2,3,$ non-minimally coupled (NMC) to gravity
with global O(3) symmetry which is broken spontaneously to U(1).
The simplest model of this kind is described by the Lagrangian
\be
{\cal L} = \sqrt{-g} \left[{ 1\over 16\pi } R + 
F(R, \phi^a\phi_a) \right] -
\sqrt{-g} \left[ {1\over 2}(\nabla \phi^a)^2 
+ V(\phi^a\phi_a) \right] \,\,\,\,,
\label{lag}
\ee
where $V(\phi^a\phi_a)$ is the 
usual Mexican hat potential
$V(\phi^a\phi_a)={\lambda\over 4}(\phi^a\phi_a-\eta^2)^2$.

Equation (\ref{lag}) shows that the introduction
of the coupling term is equivalent to consider an
effective potential
\be
V(\phi^a\phi_a)_{\rm eff}= V(\phi^a\phi_a) - F(R,\phi^a\phi_a)\,\,\,,
\ee
which explicitly depends on the matter content through $R$, and thus 
the content of ordinary matter of the galaxy 
affects the location of the minima. This feature can thus help to avoid
the scenario where the monopole configuration is universal, and opens
the possibility to recover the correlation between the masses in the
dark and ordinary matter components of the galaxy.
\bigskip

In the following we show in  detail how the non-minimal coupling
leads to the existence of bound orbits. We will focus on the case 
where $F(R,\phi^a\phi_a) = (\xi \phi^a\phi_a) R$, where $\xi$ is the 
NMC constant. 
The gravitational field equations following from
the Lagrangian (\ref{lag}) can be written as
\be
R^{\mu\nu} - {1\over 2} g^{\mu\nu}R = 8\pi G_0 T^{\mu\nu}_{\rm eff}
\ee
where
\bea\label{Teff}
T^{\mu\nu}_{\rm eff} &=&
\frac{G_{\rm eff}}{G_0} \left(4\xi T^{\mu\nu}_\xi + T^{\mu\nu}_{\rm sf}
\right) \ ,\\
\label{feq}
T^{\mu\nu}_\xi &=& \nabla^\mu(\phi^a\nabla^\nu\phi_a) - g^{\mu\nu}
\nabla_\lambda (\phi^a \nabla^\lambda\phi_a) \ ,
\label{txi} \\
T^{\mu\nu}_{\rm sf} &=& \nabla^\mu\phi^a\nabla^\nu\phi_a
- g^{\mu\nu}\left[{1\over 2}  (\nabla \phi^a)^2 + V(\phi^a\phi_a)\right] \ .
\label{tsf}
\eea
\\
The equation of motion for the scalar fields is,
\be
\Box \phi^a + 2\xi \phi^a R = {\partial V(\phi^b\phi_b)\over \partial \phi_a} \ .
\label{seq}
\ee
\\
We will only consider a metric describing spherical and static
space-times (\ref{RGMS}) and study solutions of the gravitational and
scalar fields equations describing global monopole
configurations and the resulting space-time.
Owing to the
complexity of the resulting equations, we will
perform a numerical analysis in terms of the following variables:
\bea
\nu (r) &=& {\rm ln}[N(r)] \,\,\,\,\,,\\
\label{tildenu}
\tilde \nu (r) &=& \nu(r) - \nu(0)\,\,\,\,,\\
A(r) &=& \left(1- \alpha -\frac{2G_0m(r)}{r}\right)^{-1/2}\,\,\,\,\,.
\label{AA}
\eea
where
\be
\label{alfa}
\alpha= \frac{\Delta}{1+2\xi\Delta}, \,\,\,
\Delta= 8\pi G_0 \eta^2.
\ee
The relevant Einstein equations take then the following
form
\bea
\frac{\partial m}{\partial r}
&=&  4\pi r^2 E - \frac{\alpha}{2 G_0}\,\,\,\,\,, \\
\frac{\partial \nu}{\partial r} &=&
A^2 \left\{ \frac{G_0m}{r^2} + \frac{\alpha}{2r} +
4\pi r G_0 T^{r}_{{\rm eff}\,\,r}
\right\} \,\,\,\,.
\eea
where
\be\label{E}
E= N^2 T^{tt}_{\rm eff} \,\,\,\,,
\ee
is the effective total energy density. \\
\\
On the other hand, the Klein-Gordon equation can be
written directly in terms of the energy momentum of the scalar fields:
\be
\Box \phi^a =-16\pi\xi\phi^a G_0\left(E-S\right) +
{\partial V(\phi^b\phi_b)\over \partial \phi_a} \ .
\label{seq2}
\ee
where
\be\label{S}
S= T^{i}_{{\rm eff}\,\,i}\,\,\,\,,
\ee
is the trace of the ``spatial part'' of $T^{\mu\nu}_{\rm eff}$, which
plays the role of an effective pressure.  \\
\\
In the coordinates (\ref{RGMS}) this equation reads
\bea
\frac{\partial^2 \phi^a}{\partial r^2} &=&
-\left[ \frac{2}{r} + \frac{\partial \nu}{\partial r} -
\left(1-\alpha-\frac{2 G_0 m}{r}\right)^{-1}
\left(4\pi G_0 r E -\frac{G_0 m}{r^2}-\frac{\alpha}{2r}\right)
\right] \frac{\partial\phi^a}{\partial r} \nonumber \\
& & + \left(1-\alpha-\frac{2G_0 m}{r}\right)^{-1} \left[
{\partial V(\phi^b\phi_b)\over \partial \phi_a}
-16\pi\xi\phi^a G_0\left(E-S\right) \right] \nonumber \\
& & - \frac{1}{r^2} \left(1-\alpha-\frac{2 G_0 m}{r}\right)^{-1}
\left[\frac{\partial^2 \phi^a}{\partial \theta^2} +
\frac{\cos\theta}{\sin\theta} \frac{\partial \phi^a}{\partial \theta} +
\frac{1}{\sin^2\theta} \frac{\partial^2 \phi^a}{\partial \varphi^2}
\right] \,\,\,\,\,.
\eea
The ansatz for a monopole configuration is
\be\label{conf}
\phi^a = \eta f(r) \frac{x^a}{r},
\ee
with $x^a x^a = r^2$, so that a monopole
solution is found if $f \rightarrow 1$ at spatial infinity
(i.e., $||\phi^a||\rightarrow \eta$).  \\
\\
It is clear from the Eqs.
(\ref{Teff}--\ref{tsf}),
that the intermediary variables $E$ and $S$
[see Eqs.(\ref{E}), (\ref{S})] involve second
order derivatives of the scalar field.
However, we can eliminate such a terms from the gravitational field
equations with the help of the ansatz for the monopole field and of
the Klein-Gordon equation, and obtain
``sources''  containing at most first order derivatives
of the scalar field.
We also introduce the following dimensionless quantities
\\
\bea
\tilde r &:=& r\cdot \eta\lambda^{1/2}  \,\,\,\,,\\
\tilde m &:=& m\cdot G_0 \eta\lambda^{1/2} \,\,\,\,,\\
\label{rotilde}
\tilde \rho &:=& \rho\cdot \frac{G_0}{\eta^{2}\lambda}  \,\,\,\,,\\
\tilde p &:=& p\cdot \frac{G_0}{\eta^{2}\lambda} \,\,\,\,,\\
\tilde \phi^a &:=& \frac{\phi^a}{\eta}  \,\,\,\,,\\
\frac{\phi^a}{\eta} &:=& f(r) \frac{x^a}{r}  \,\,\,\,,\\
\tilde V(\tilde \phi^a\phi_a) &:=& V(\phi^a\phi_a)\cdot \frac{G_0}{\eta^{2}\lambda}
\,\,\,\,,\\
\Delta &:=& 8\pi G_0 \eta^2  \,\,\,\,,\\
\alpha &:=& \frac{\Delta}{1+2\xi\Delta}  \,\,\,\,,\\
\tilde G_{\rm eff} &:=&\frac{1}{1+ 2\xi \Delta f^2}\,\,\,\,,
\eea
\\
then the final form of the equations to be analyzed numerically is :
\\
\bea
\label{massu}
\partial_{\tilde r} \tilde m &=&  4\pi \tilde r^2 \tilde E -
\frac{\alpha}{2}\,\,\,\,\,,\\
\label{lapsefi}
\partial_{\tilde r}\tilde \nu  &=& \frac{A^2}{1+2\xi\Delta
\tilde r f (\partial_{\tilde r}f) \tilde G_{\rm eff} }
\left\{ \frac{\alpha}{2\tilde r} + \frac{\tilde m}{\tilde r^2} +
\frac{\Delta}{2} \tilde r \tilde G_{\rm eff}\left[
\frac{1}{2A^2}(\partial_{\tilde r}f)^2 - \frac{(f^2-1)^2}{4}
 - \frac{f^2}{\tilde r^2} - \frac{8\xi f (\partial_{\tilde r}f)}
{\tilde rA^2}\right]  \right\}, \,\,\,\, \\
\label{scalarfi}
\partial_{\tilde r\tilde r} f &=&
-\left[ \frac{2}{\tilde r} + \partial_{\tilde r} \tilde \nu -
\left(1-\alpha-\frac{2\tilde m}{\tilde r}\right)^{-1}
\left(4\pi \tilde r \tilde E - \frac{\alpha}{2\tilde r}
-\frac{\tilde m}{\tilde r^2}\right)
\right] (\partial_{\tilde r}f) \nonumber \\
& & + \left(1-\alpha-\frac{2\tilde m}{\tilde r}\right)^{-1} \left[
f(f^2-1) + \frac{2f}{\tilde r^2}
-16\pi\xi f \left(\tilde E-\tilde S\right) \right], \,\,\,\,\,
\eea
where
\bea\label{e-s}
\tilde E-\tilde S &=&
\frac{\Delta\tilde G_{\rm eff}}{8\pi(1+ 24\Delta\xi^2 f^2 \tilde G_{\rm eff})}
\left[ \left(\frac{1}{A^2}(\partial_{\tilde r} f)^2 + \frac{2f^2}{\tilde r^2}
\right) (1 + 12\xi) + (f^2-1)^2
+ 12\xi f^2 (f^2-1) \right], \,\,\,\,\\
\tilde E &=&
\frac{\Delta\tilde G_{\rm eff}}
{8\pi(1+ 24\Delta\xi^2 f^2 \tilde G_{\rm eff})}
\left[-\frac{4\xi f(\partial_{\tilde r} f)(\partial_{\tilde r}\tilde \nu)}{A^2}
\left( 1+ 24\Delta \xi^2 f^2 \tilde G_{\rm eff} \right)
+ 4\xi f^2 (f^2-1) \right.  \nonumber \\
\label{Eu}
&& \left.
\quad \quad \quad \quad \quad \quad \quad
\quad \quad \quad \quad
+ \left(
\frac{1}{2A^2}(\partial_{\tilde r} f)^2 +
\frac{f^2}{\tilde r^2} \right)
\left(1+8\xi +
8\Delta\xi^2 f^2 \tilde G_{\rm eff}\right) \right. \nonumber \\
&& \left.   \quad \quad \quad \quad \quad \quad \quad
\quad \quad \quad \quad
+ \frac{(f^2-1)^2}{4} \left(1 - 8\Delta\xi^2 f^2
\tilde G_{\rm eff} \right) \right].
\eea
Here $\tilde E$ and $\tilde S$ are dimensionless as in
Eq. (\ref{rotilde}).

We note now that the sources of the differential equations contain only
first order derivatives of the field variables and are thus suitable for
numerical integration with a Runge-Kutta algorithm.

\subsection{Asymptotic expansions and boundary conditions}
\bigskip

Let us discuss the asymptotic behavior of global
monopoles at the origin and spatial infinity in order to
find the boundary conditions for the numerical
integration.
The regularity condition at $r=0$ on the metric
requires
\bea\label{cfm}
\tilde m(0)= 0 \ , \partial_{\tilde r}\tilde m(0)= -\alpha/2\,.
\eea
The boundary condition on $\tilde \nu(r)$ is by definition
\bea\label{cfn}
\tilde \nu (0) \equiv 0 \,\,.
\eea
The boundary condition on the scalar
field at $r=0$ is (false vacuum),
\bea\label{cfce}
f(0) &=& 0  \,\,.
\eea
Then one finds the
expansions of the functions
$\tilde m(\tilde r)$, $f(\tilde r)$ and
$\tilde\nu(\tilde r)$, at $r=0$,
\\
\bea
\tilde m(\tilde r)= -\frac{\alpha}{2}\tilde r + 
\frac{\Delta}{12} \left[\frac{1}{2} + 3(8\xi + 1) f_c^2\right] \tilde r^3 +
O(\tilde r^{4})  \,\,\,,
\eea
\bea
f(\tilde r)= f_{c} \tilde r + O(\tilde r^{3}),
\eea
\bea
\tilde \nu(\tilde r)= -\frac{\Delta}{24}(1 + 24\xi)f_c^2 \tilde r^2 + 
O(\tilde r^{3}),
\eea
\\
where $f_{c}$ is determinated by
the boundary conditions at spatial infinity.
\\
Let define
$M = \tilde m(\infty)$. As we mentioned, we consider monopole
configurations (i.e., $f \rightarrow 1$ at spatial infinity: true vacuum). 
Then we have the asymptotic expansions at spatial infinity,
\bea
\tilde m(\tilde r)= M +
\frac{\alpha}{2\tilde r \Delta}
\left[ \frac{\alpha^2}{\Delta}+
\frac{8\xi(1-\alpha)}{\alpha} \right] +
O(\tilde r^{-2}) \,\,,
\eea
\bea
f(\tilde r)= 1 - \frac{1}{(1+2\xi\Delta)\tilde r^2}
+ O(\tilde r^{-4}) \,\,,
\eea
\bea
\tilde \nu(\tilde r)= \tilde \nu(\infty) -
\frac{M}{(1-\alpha)\tilde r} +
O(\tilde r^{-2}) \,\,.
\eea
\\
We will impose for the asymptotic behavior of the metric,
the standard asymptotically-flat-but-for-a-deficit-angle
$\alpha$ spacetime (S.A.F.D.A $\alpha$) (see \cite{ulises})
and therefore the boundary condition on the function
$\nu(\tilde r)$ at spatial infinity is
\bea\label{cfinf}
\nu(\infty) = \frac{1}{2} \ln (1 - \alpha) \,\,.
\eea

The integration of the equations is performed by specifying
the regularity condition on the scalar field at $r=0$,
\bea
\partial_{\tilde r} f (0) &=& f_c
\,\,\,\,.
\eea
The value $f_c$ cannot be arbitrary,
but must be so that $f$ satisfy the appropriate boundary conditions at
spatial infinity. This is enforced by the use of a standard shooting
method \cite{recipes}.

We can compute the solution by integrating the equations in one step,
from $r=0$ to a radius which is chosen
conveniently so that $f = 1$ with a certain
degree of approximation.
The physical lapse $N(r)= e^{\nu}$ at $r=0$
is calculated at the end of
the numerical integration from (\ref{tildenu}) and
(\ref{cfinf})
\be \nu (0)= \frac{1}{2} \ln (1 - \alpha)-
\tilde \nu_\infty \,\,\,\,,
\ee
where the value $\tilde \nu_\infty$ is obtained
from the numerical integration. This ensures that at spatial infinity 
we recover the S.A.F.D.A $\alpha$ spacetime.

\bigskip
We note that the ADM
mass of the configurations of global monopoles,
(see \cite{ulises} for a rigorous 
definition of the ADM mass for the case of spacetimes with a deficit
angle) can be easily
computed from the integral
\be
M_{\rm ADM\alpha} = M = {\rm lim}_{r\rightarrow\infty} \,\,
m (r) = \int_{0}^{\infty}
\left(4\pi r^2 E(r) - \frac{\alpha}{2}\right) dr \,\,\,\,.
\ee

\subsection{Results.}
\bigskip

In order to obtain static configurations, we must impose
the condition $0 < \alpha < 1$ for the deficit angle.
The behavior of the deficit angle $\alpha$ when $\Delta \sim 0$ is,
\\
\bea
\alpha \sim \Delta - 2\xi\Delta^2 + O(\Delta^3).
\eea

\bigskip

\begin{figure*}
\vspace{1cm}
\psfig{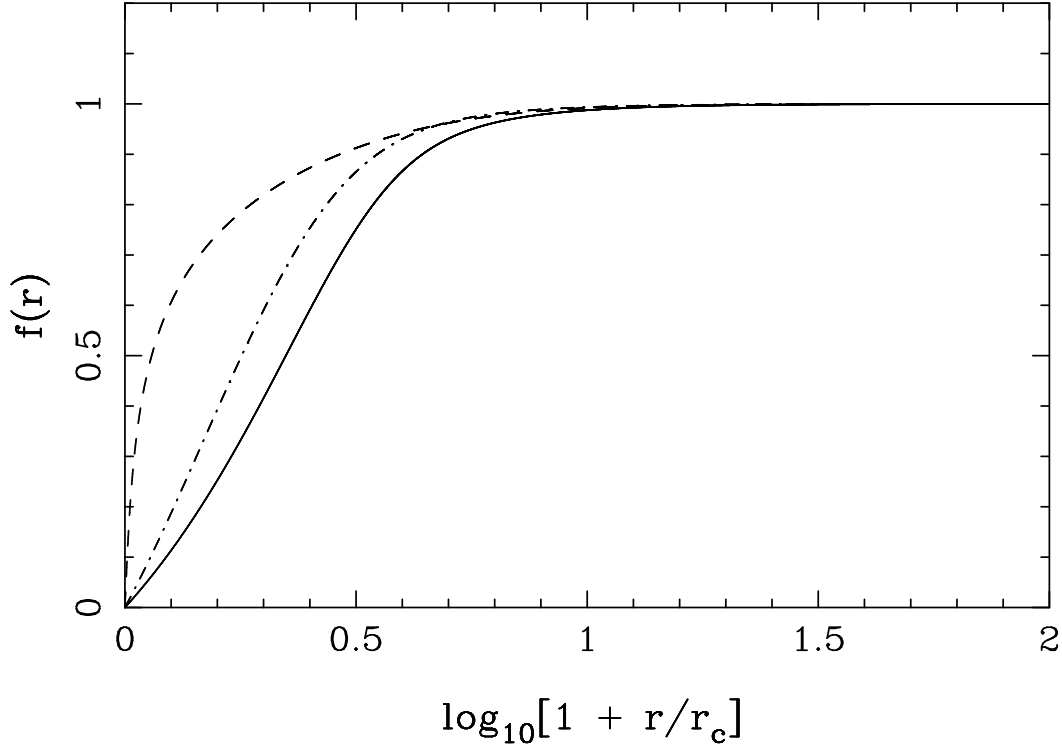}
\hspace*{-2.5in} %%\vskip 2cm
\caption[]{\label{f:scalar}
The figure depicts the global monopole field $f(r)$
for $\xi=0$, $\alpha=0.795$ (solid line),
$\xi=2$, $\alpha=0.1$ (dashed line), and 
$\xi=0.3$, $\alpha=0.79$ (dash-dotted line).
Here $r_c \equiv (\eta \lambda^{1/2})^{-1}$.}
\end{figure*}

\begin{figure*}
\vspace{1cm}
\psfig{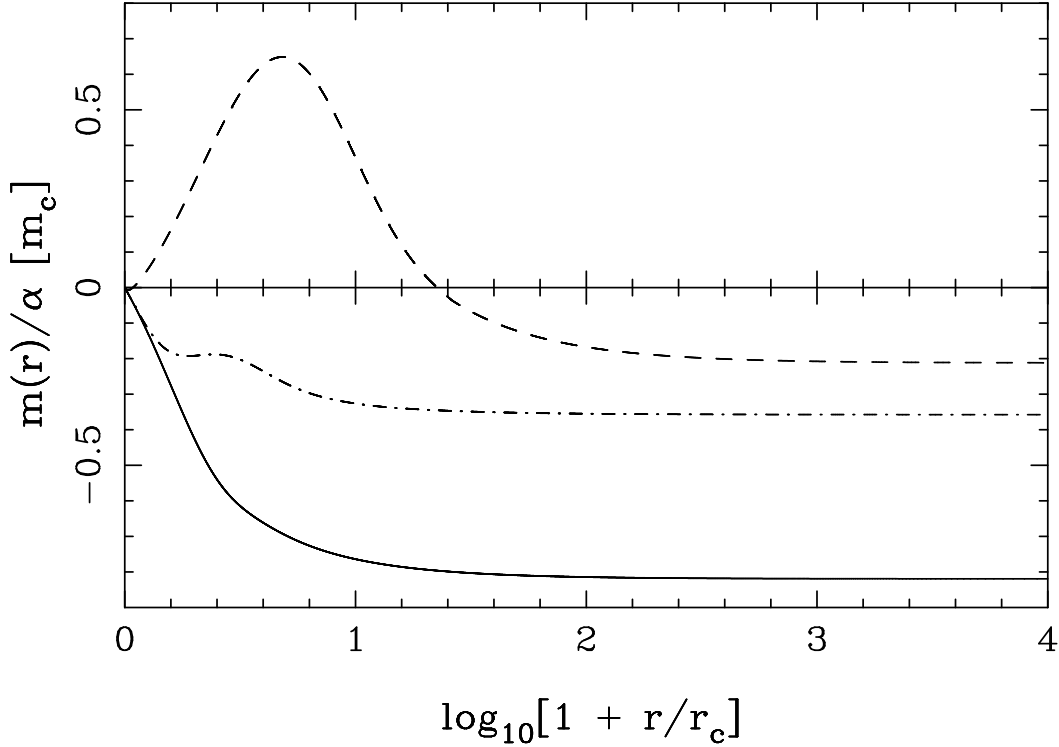}
\hspace*{-2.5in}
\caption[]{\label{f:confmasa}
Mass profile $m(r)$ for $\xi=0$, $\alpha=0.795$ (solid line),
$\xi=2$, $\alpha=0.1$ (dashed line), and 
$\xi=0.3$, $\alpha=0.79$ (dash-dotted line).
Asymptotically this quantity
provides the ADM mass of the configuration.
Here $m_c \equiv (G_0 \eta \lambda^{1/2})^{-1}$.}
\end{figure*}

\begin{figure*}[h]
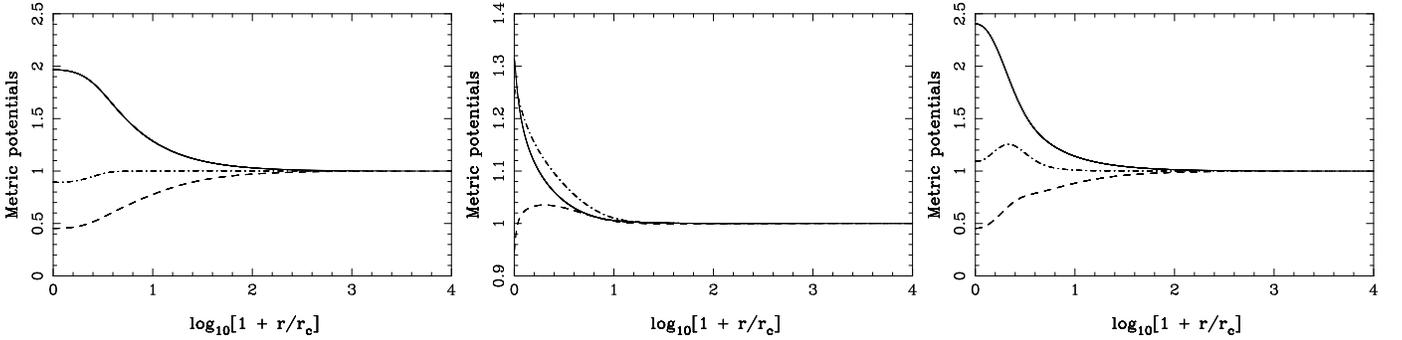

\centerline{
\psfig{figure=lapsemasa0.795.ps,angle=-90,width=6cm}
\psfig{figure=lapsemasa2.1.ps,angle=-90,width=6cm} % \hspace*{-2.5cm}
\psfig{figure=lapsemasa.3.79.ps,angle=-90,width=6cm} }
\vspace*{0.5cm}
\caption[]{\label{f:metricpot}
Metric potentials $N(r)/(1-\alpha)^{1/2}$ (solid lines),
$A(r)(1-\alpha)^{1/2}$ (dashed lines), and the product $AN$
(dash-dotted lines) for three different configurations. 
The left panel corresponds to $\xi=0$, $\alpha=0.795$, the 
middle panel to $\xi=2$, $\alpha=0.1$, and the right panel 
to $\xi=0.3$, $\alpha=0.79$. Note that $AN$ aproaches to unit 
``outside'' the monopole core, while the metric potentials 
$N$ and $A$ tend to the asimptotically-flat-but-for-a-deficit-angle 
values far from the origin.}
\end{figure*}

{\bf Case $\xi > 1/2$.} The range allowed
for the breaking scale is $\Delta \in [0, \infty)$ and
for the deficit angle $\alpha \in [0, \frac{1}{2\xi})$.
This is consistent with the condition $\alpha < 1$.
\\
Figures \ref{f:scalar}$-$\ref{f:metricpot}, shows
numerical solutions for the case $\xi=2$ satisfying the required boundary
conditions.

{\bf Case $\xi \leq 1/2$.} The range permitted
for the breaking scale is $\Delta \in [0, 1/(1-2\xi\alpha)]$ and
for the deficit angle $\alpha \in [0, 1]$.
In this case the static configurations cease to exist when
$\Delta > 1/(1-2\xi\alpha)$.
\\
Figures \ref{f:scalar}$-$\ref{f:metricpot}, shows
numerical solutions for the cases $\xi=0.3$ and $\xi=0$ 
(minimal coupling case) satisfying the required boundary
conditions.

\begin{figure*}
\vspace{1cm}
\psfig{figure= 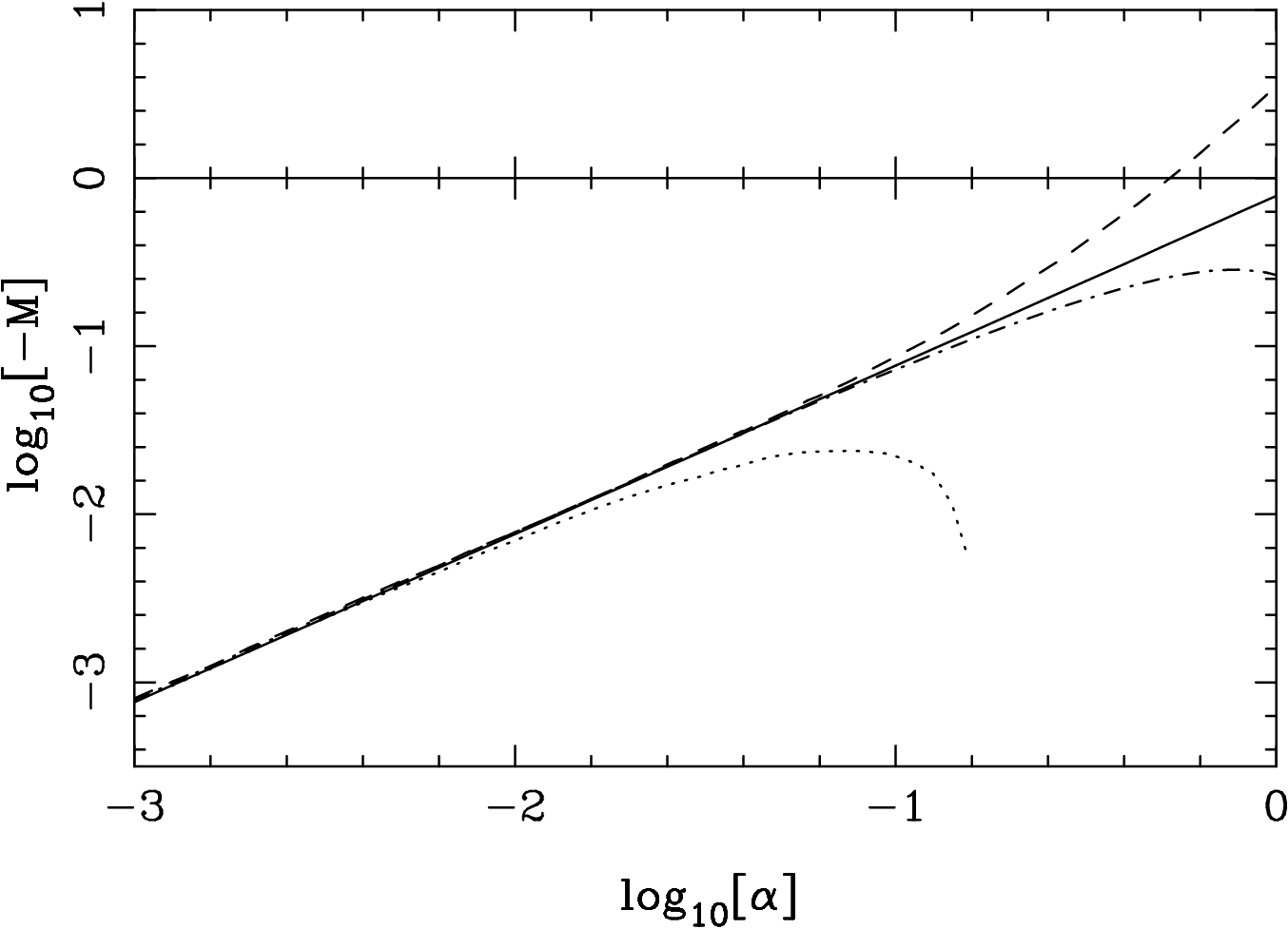,angle=-90,width=5.5in}
\hspace*{-2.5in}
\caption[]{\label{f:logmasa}
ADM mass of configurations for different values of the
deficit angle for $\xi=0$ (solid line),
$\xi=-2$ (dashed line),  
$\xi=0.3$ (dash-dotted line) and
$\xi=2$ (dotted line).}
\end{figure*}

\begin{figure*}
\vspace{1cm}
\psfig{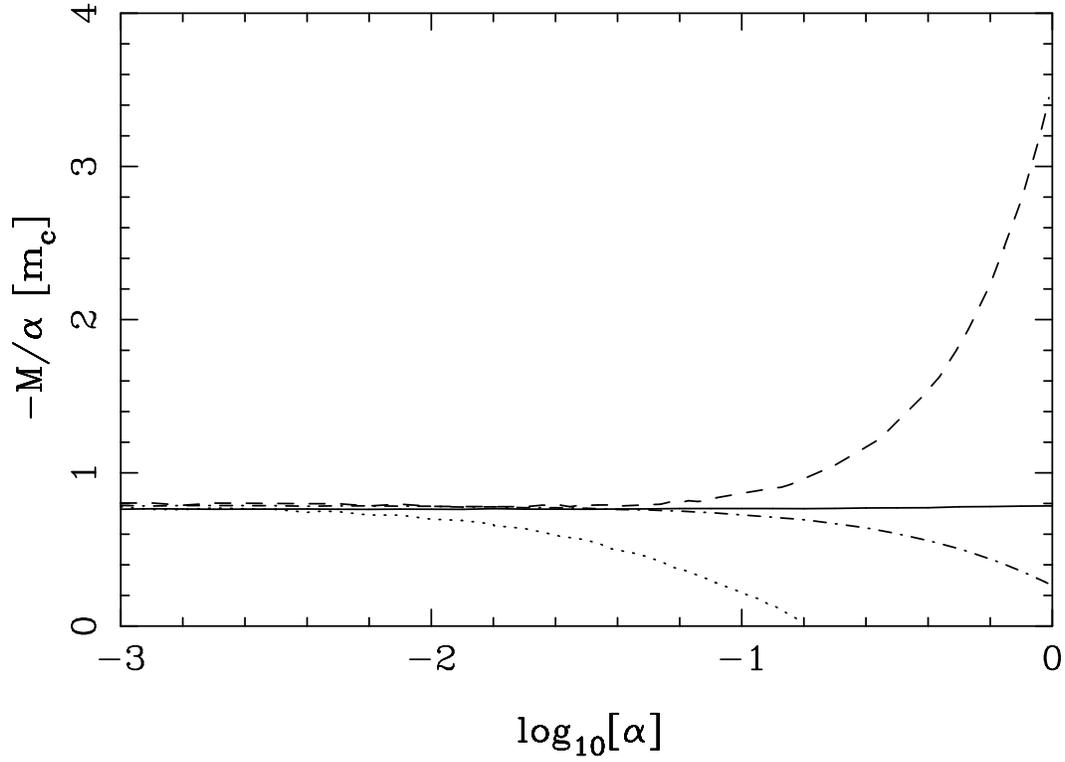}
\hspace*{-2.5in}
\caption[]{\label{f:admmassalfa}
ADM mass-deficit angle rate of configurations for different 
values of the deficit angle for $\xi=0$ (solid line),
$\xi=-2$ (dashed line), $\xi=0.3$ (dash-dotted line), and
$\xi=2$ (dotted line).}
\end{figure*}

\begin{figure*}
\vspace{1cm}
\psfig{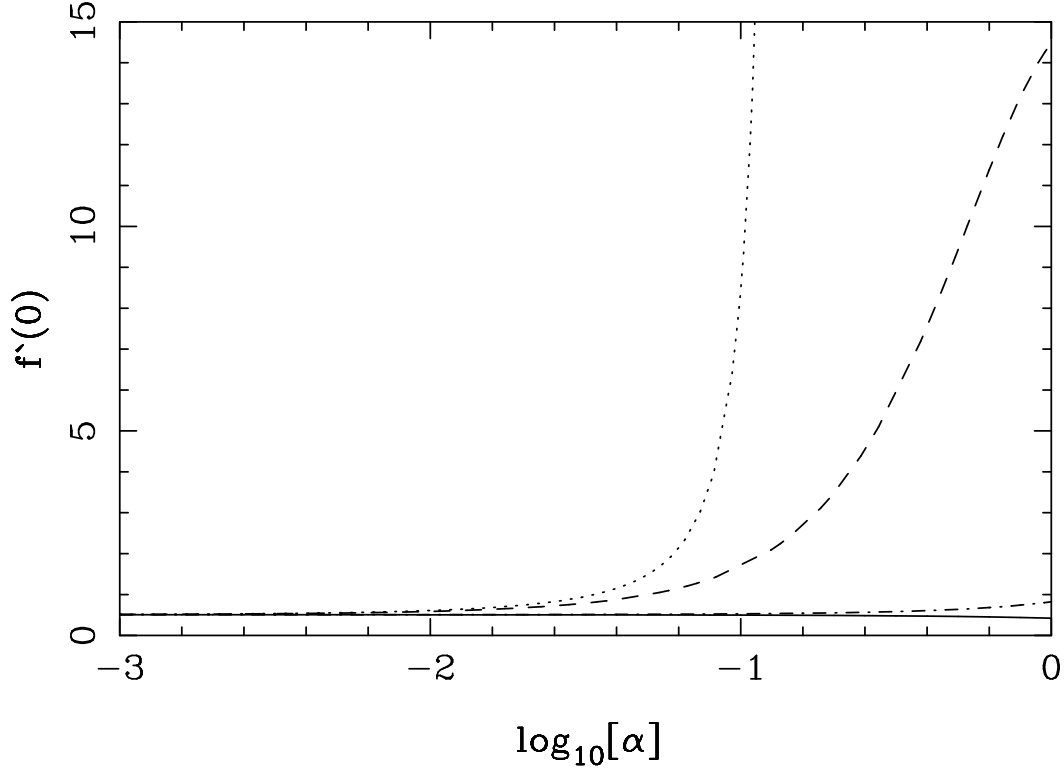}
\hspace*{-2.5in}
\caption[]{\label{f:deri}
Derivative of the global monopole at $r=0$ for different 
values of the deficit angle for $\xi=0$ (solid line),
$\xi=-2$ (dashed line), $\xi=0.3$ (dash-dotted line), and
$\xi=2$ (dotted line).}
\end{figure*}

In the minimal coupling case Harari \& Loust\'o \cite{harari} estimated
analytically the mass of the monopole as $M\sim -2\alpha/3$, and
the size of the core $\tilde\delta\sim 2$. 
Then numerically showed that $M\sim -0.75\alpha$. That is,
they showed that the ratio $M/\alpha$ is practically
insensitive to $\alpha$ (see figs.\ref{f:logmasa},\ref{f:admmassalfa}).  \\
\\
In the nonminimal coupling case, for a given $\xi$ the situation is
similar in that the ratio $M/\alpha$ is practically
insensitive to $\alpha$ for $\alpha \leq 0.01$. Actually, in this 
range of small $\alpha$, that ratio depends weakly on the value of $\xi$. 
However, for $\alpha > 0.01$ the ratio depends strongly on $\xi$ and 
$\alpha$ [cf. figs.\ref{f:logmasa},\ref{f:admmassalfa}].
This can be seen by performing an analysis similar to the one of
\cite{harari} but assuming the following approximation
\\
\be
f = \left\{ \begin{array}{lll}
                f_c \tilde r & \mbox{if} & \tilde r < \tilde \delta \\
                1- \frac{1}{(1+2\xi\Delta)\tilde r^2}& \mbox{if} &
\tilde r > \tilde \delta
              \end{array} 
\right.  
\ee

and for the function $\tilde m(\tilde r)$
\\
\be
\tilde m = \left\{ \begin{array}{lll}
                -\frac{\alpha}{2}\tilde r + 
\frac{\Delta}{24} \tilde r^3 
& \mbox{if} & \tilde r < \tilde \delta \\
                M & \mbox{if} &
\tilde r > \tilde \delta
              \end{array}
       \right.  
\ee

matching continuously the function $\tilde m$ and its derivative at
$\tilde r = \tilde \delta$, we obtain,
\\
\begin{eqnarray}
M &=& - \frac{2}{3} \frac{\Delta}{(1 + 2\xi \Delta)^{3/2}} =
- \frac{2}{3} \alpha (1 + 2\xi \Delta)^{-1/2} \,\,\,,
\\
\tilde \delta &=& \frac{2}{(1 + 2\xi \Delta)^{1/2}} = 
2 (1 - 2\xi \alpha)^{1/2} 
\end{eqnarray}
\\
where we have used the relation 
$\Delta = \frac{\alpha}{(1 - 2\xi \alpha)}$. 
For the case $\xi = 0$, we reproduce for $M$ and $\tilde \delta$ 
the values estimated analytically
by Harari \& Loust\'o \cite{harari}. We can see that in the case
$\xi > 1/2$, we have $\Delta \in [0, \infty)$ and 
$\alpha \in [0, \frac{1}{2\xi})$ and therefore $M \rightarrow 0$
when $\alpha \rightarrow \frac{1}{2\xi}$, these is confirmed numerically 
for the value $\xi=2$[cf. fig.\ref{f:admmassalfa}]. 
Now we match continuously the function $f$ at $\tilde r = \tilde \delta$
and we obtain
\\
\begin{eqnarray}
f_c &=& \frac{2}{3\sqrt 3} (1 + 2\xi \Delta)^{1/2} = 
\frac{2}{3\sqrt 3} (1 - 2\xi \alpha)^{-1/2}  \,\,\,, \\
\end{eqnarray}
\\
then when $\alpha \rightarrow \frac{1}{2\xi}$ we have that
the value of $f_c$ diverges, 
this is checked out numerically for the value $\xi=2$
[cf. fig.\ref{f:deri}].
 
\subsection{Geodesic motion in the spacetime of a global monopole.}
\begin{figure*}[h]
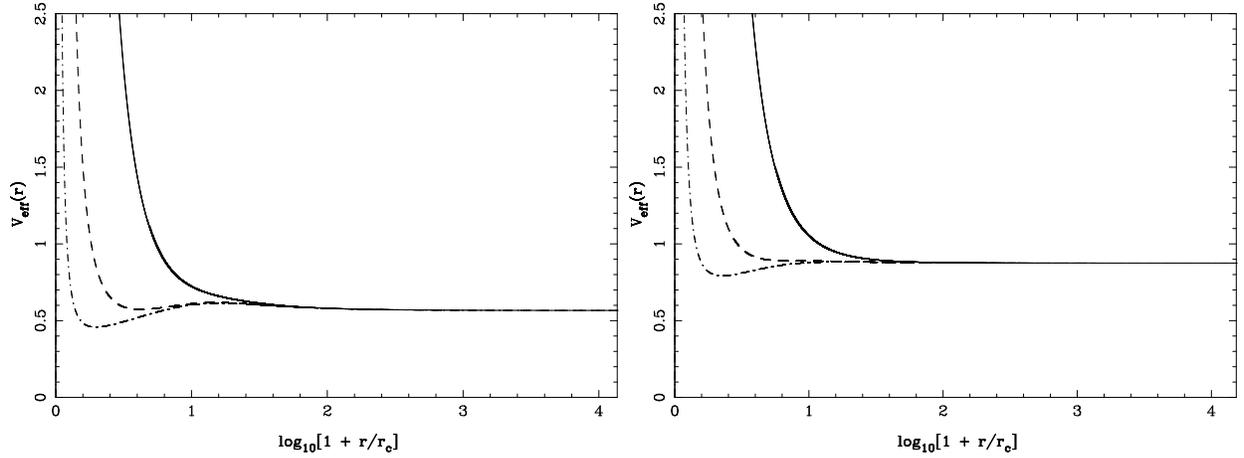

\centerline{
\psfig{file= poteff-2.ps,angle=-90,width=8.1cm} % \hspace*{-2.5cm}
\psfig{file= poteff-21.ps,angle=-90,width=8.1cm}}
\vspace*{0.5cm}
\caption[]{\label{f:velo1}
Functional dependence of the effective potential
$V_{eff}$ vs $\tilde r$ for the case
$\xi=-2$, $\alpha=0.43$ (left panel)
and $\alpha=0.125$ (right panel).
For each configuration we show three values of the
angular momentum: $L=4$ (solid line),
$L=1$ (dashed line) and $L=0.3$ (dash-dotted line).
Here $r_c \equiv (\eta\lambda^{1/2})^{-1}$.}
\end{figure*}

\begin{figure*} \vspace{1cm}
\psfig{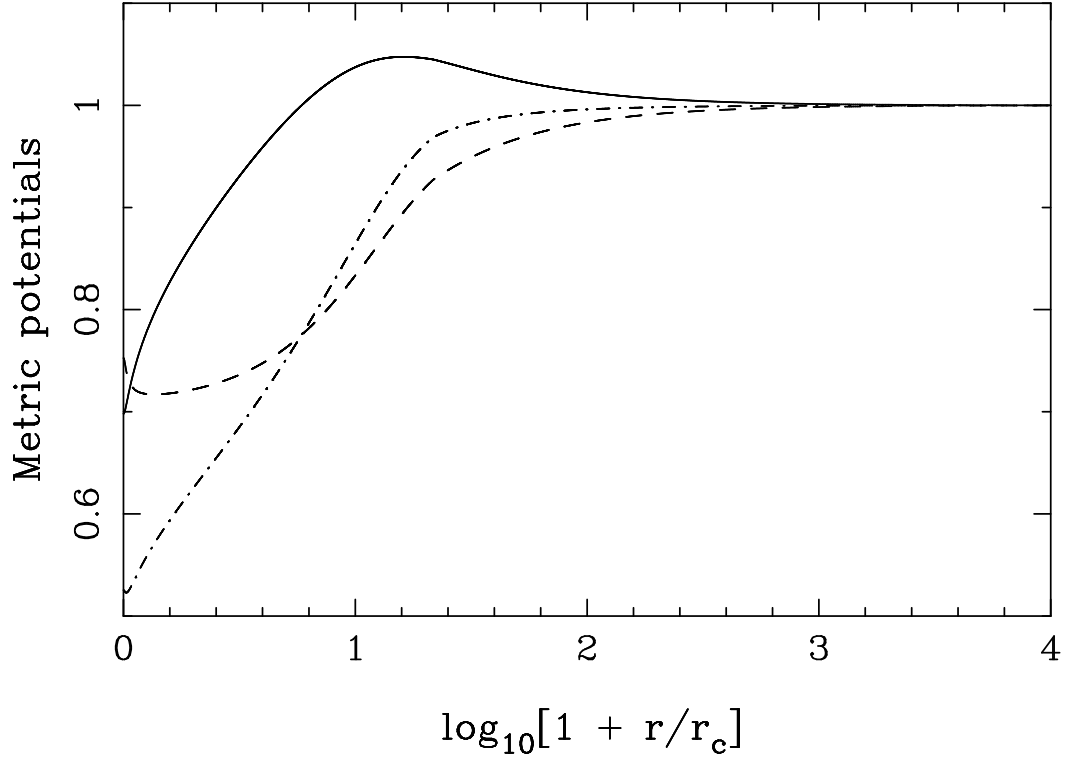} \hspace*{-2.5in}
\caption[]{\label{f:metric-2} Same as fig.\ref{f:metricpot}
for $\xi=-2$, $\alpha= 0.43$. }
\end{figure*}

\begin{figure*}[h]
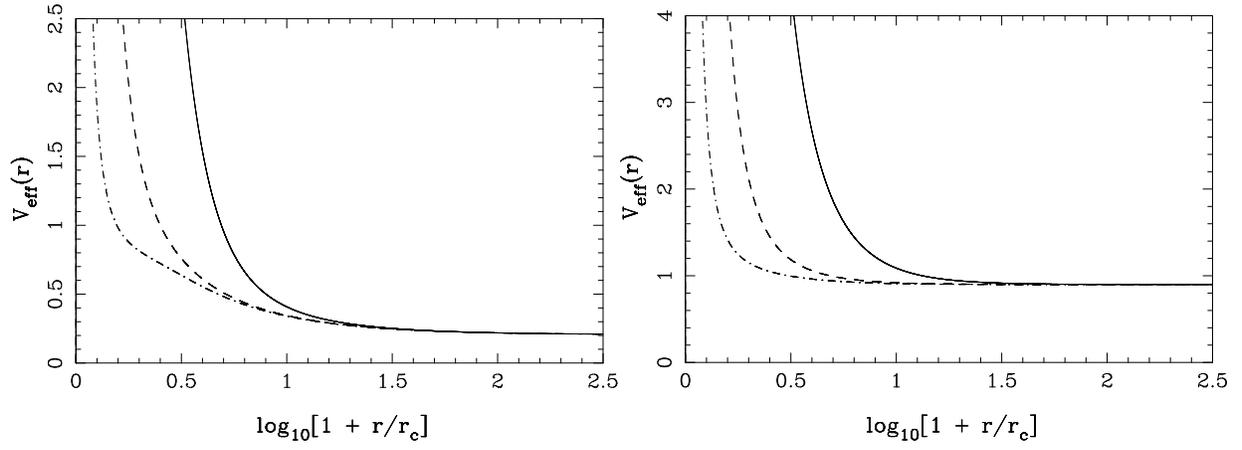

\centerline{
\psfig{figure=Veff0.795.ps,angle=-90,width=8.1cm}% \hspace*{-2.5cm}
\psfig{figure=Veff2.1.ps,angle=-90,width=8.1cm} }
\vspace*{0.5cm}
\caption[]{\label{f:velo2}
 Same as fig.\ref{f:velo1}
 for $\xi=0$, $\alpha= 0.795$ (left panel) and 
 for $\xi=2$, $\alpha= 0.1$ (right panel).}
\end{figure*}

\bigskip
In order to analyze the geodesic motion
of test particles in the spacetime generated by a global monopole
we consider Eq.(\ref{veffe}). In this case $N_D= (1-\alpha)^{1/2}$
and the remaining gravitational potentials are given numerically.
Figure \ref{f:velo1} shows the effective potential
for $\xi=-2$ and for different values of the other parameters;
here we note the existence of a potential well and a non trivial 
minimum and thus the existence of stable circular orbits.  \\
\\
For $\xi>0$ the effective potential (\ref{veffe}) does not exhibit
maxima or minima.
An heuristic analysis that helps to understand the appearance of
extrema of $V_{eff}$ when $\xi <0$ and their absence when 
$\xi>0$, is the following. We have that,
$\partial_r V_{eff}=  2N^2(r) ( -L^2/r^3 + L^2\partial_r \nu/r^2 +
\partial_r \nu)$. So for $\partial_r V_{eff}=0$ and then to find 
extrema, notably minima, it is necessary that
$\partial_r \nu > 0$ (or equivalently $\partial_r N>0$). This means
that $N(r)$ should have an increasing behavior as a function of 
$r$ at least in some region away from the origin. Eq. (\ref{lapsefi})
provides the sign for the slope of $N$. Near the origin $f\sim f_c r$
and then $\partial_r \nu\sim -(\Delta/12)(1 + 24\xi) rf_c^2 + O(r^3)$. 
For $\partial_r \nu> 0$, $\xi$ must be negative
enough so that the coefficient $(1 + 24\xi)$ is negative 
[cf. Fig.\ref{f:metric-2}].
On the other hand, for $\xi\geq 0$, $\partial_r 
\nu < 0$ and therefore $V_{eff}$ has no minima (no bound orbits)
[cf. Figs.\ref{f:metricpot}, \ref{f:velo2}].
This heuristic argument is confirmed by the rigorous numerical analysis
from which the critical value $\xi_{\rm crit}$ that allows
the existence of bound orbits is found to be 
$\xi_{\rm crit} \lesssim -0.15$.

Moreover, for $L=0$ which corresponds to $V_{eff}\equiv N^2$, a
peculiar situation occurs in the cases where $V_{eff}$ has extrema,
notably a maxima (e.g., for $\xi=-2$). The maxima and minima of $V_{eff}$
will correspond to the locus of unstable and stable
stationary points where test particles are static (i.e., 
where particles do not feel any gravitational field). As seen from
fig.\ref{f:metric-2}, we appreciate that at the origin $r=0$, test
particles can be at rest in stable equilibrium, while at
$r_{\rm max}$ where $V_{eff}$ is maximum test particles can be
in unstable static equilibrium. In other
words, $r_{\rm max}$ separates two regions: one attractive and
other repulsive [cf. the $N(r)$ profile from Fig.\ref{f:metric-2}]. 
This strange behavior
contrast dramatically from the cases of
``conventional'' gravitational sources like stars, 
planets, etc. where test particles are always attracted
towards the source and where there are ``no trivial'' points at 
which they can remain static. In the case of minimally coupled 
global monopoles test particles are always repelled. 

\begin{figure*}[h]
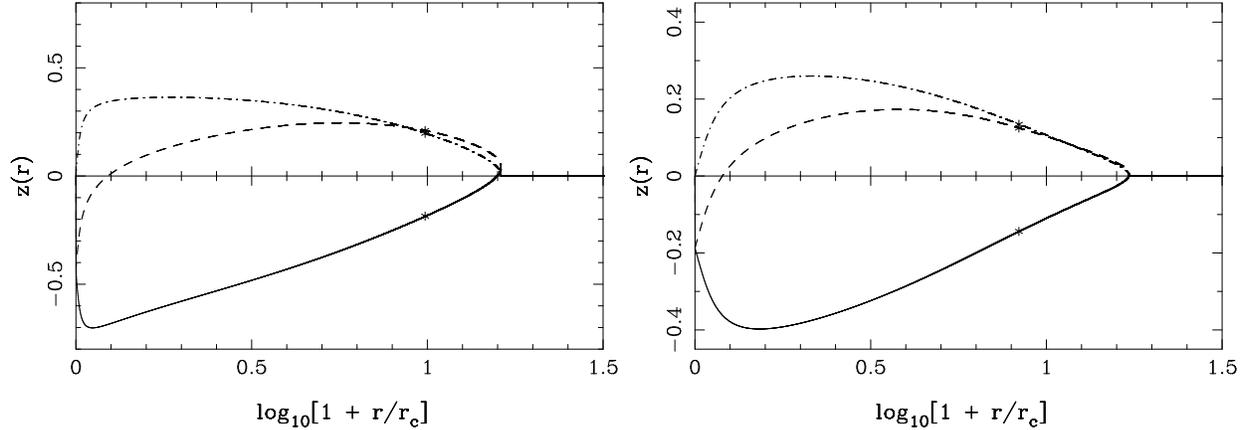

\centerline{
\psfig{file= rs.ps,angle=-90,width=8.1cm} % \hspace*{-3cm}
\psfig{file= rs1.ps,angle=-90,width=8.1cm}}
\vspace*{0.5cm}
\caption[]{\label{f:rs}
Functional dependence of $z_+$ (dashed line),
$z_-$ (solid line) and $z_D$ (dash-dotted line)
vs $\tilde r$ for circular orbits in the case
$\xi=-2$, $\alpha=0.43$ (left panel) and $\alpha=0.125$
(right panel). The asterisk depicts the location of the
radius beyond which the stable circular orbits cease to exist
($r \sim 9 \,r_c$).}
\end{figure*}

\begin{figure*} \vspace{1cm}
\psfig{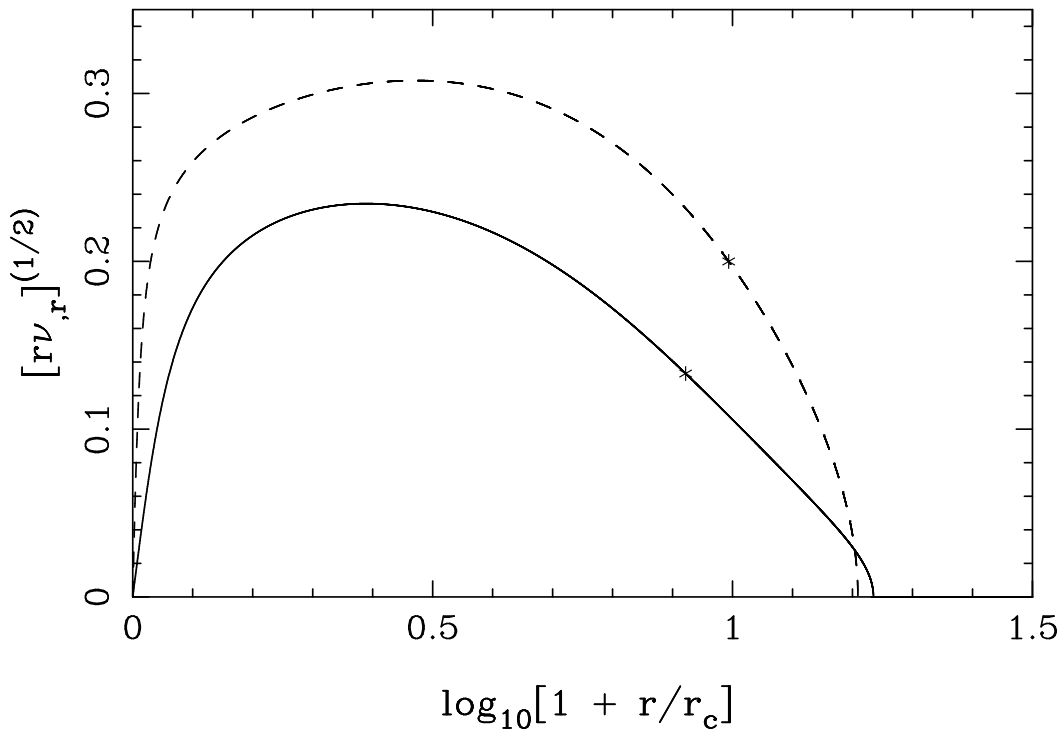} \hspace*{-2.5in}
\caption[]{\label{f:vtang} Tangential velocity
$v= (r\partial_rN/N)^{1/2}$ in units of $c$ for 
$\xi=-2$, $\alpha=0.125$ (solid line) and 
$\alpha=0.43$ (dashed line).
The asterisk depicts the location of the
radius beyond which the stable circular orbits cease to exist.}
\end{figure*}

Since bound orbit exists in
the spacetime generated by nonminimal global monopole with suitable
$\xi$, we can compute the corresponding shift $z_D$
from Eq.(\ref{rsm}) using the numerical solutions for $N$ and
compare with the RC of spiral galaxies.
Figure \ref{f:rs} depicts $z_+$ (dashed line),
$z_-$ (solid line) and $z_D$ (dash-dotted line)
as functions of $\tilde r$ for the cases $\alpha = 0.43$
(left panel) and $\alpha = 0.125$ (right panel)
respectively. We note that even for this very simple model
the figures that would correspond to the rotation curves contain
a relatively ``flat region'' within the values of $r$ corresponding
to stable orbits (i.e. the behavior of $z_D$ near its maximum). \\
\\
From these figures it is interesting to note that $z_+$ that in principle
is associated with a blue-shift, does not always correspond to a blue-shift,
since there is a value $r_b$ that separates $z_+$ of being positive
(blue-shift) or negative (red-shift). This is easy to understand since
the value $z_+$ arise from a competition between the gravitational red-shift
and the kinematical blue-shift. For slow particles 
(i.e, particles orbiting at ``small'' $r$) moving in the same direction  
as the emitted light, 
the gravitational barrier dominates over
the positive contribution of the kinematical effects, and thus the
frequency of emitted light has an overall attenuation. At
orbits with radius $r_b$, particles are fast enough for the kinematical
blue-shift to cancel the gravitational red-shift resulting
in $z_+=0$. For example, from (\ref{ztf}) and for $v<<1$,
it turns that $z_+ < 0$ if
$v< 1 - N(r)/(1-\alpha)^{1/2}$. We note that in the non-minimal coupling case
$(1-\alpha)^{1/2} > N(r)$ in the regions where bound orbits exist.
For larger values of $r$, $z_+$ reaches a 
maximum and then starts decreasing since 
the gravitational barrier becomes larger (see Fig.\ref{f:velo1})
while the tangential velocity $v$ [$v= (r\partial_r N/N)^{1/2})$] 
becomes smaller until reaching zero at 
the radius $r_{\rm max}$ where $N$ is maximum (see Fig.\ref{f:vtang}).  \\
\\
Concerning $z_-$, this quantity is generically 
negative (i.e., it corresponds to a true red-shift). For instance,
when $v<<1$, then $z_- \approx 1- (1-\alpha)^{1/2}(1+v)/N$ so
$z_-< 0$ if $(1-\alpha)^{1/2}(1+v)/N > 1$. This condition
holds in most of the region of bound orbits since then
$N< (1-\alpha)^{1/2}$ and $v\neq 0$ (see Figs.\ref{f:metric-2}).
However, moving away
from the origin $v\rightarrow 0$ and $N$ grows to a maximum value where
$v=0$ and $z_-= z_+ = 1- (1-\alpha)^{1/2}/N$ which can 
be positive. For the cases (values of $\xi$) giving rise to bound orbits, 
it seems that $N$ always has a global maximum and then 
$N_{\rm max} >  (1-\alpha)^{1/2}$.
Therefore $z_\pm>0$ at $r_{\rm max}$. In fact the region of $r$ where
$z_-\geq 0$ is very narrow and corresponds to 
$(1-\alpha)^{1/2}(1+v)/N \leq 1$
[practically unseen at the scales of Fig.(\ref{f:rs}); this 
corresponds to orbits of small angular momentum]. \\

To conclude this section, we mention that although the nonminimal 
global monopole model can repair the two main objections posed on the 
minimal model (namely, the bound orbit and the correlation between 
luminous and dark matter problems), there are still several improvements to 
perform in order that the quantitative predictions of this model 
fit reasonably well with the astrophysical data. 
Therefore, it is still very premature for any claim on this 
model as a realistic candidate for explaining the galactic dark matter 
and the corresponding rotation curves. 
The fact that the model is fully consistent in 
what regards the mathematical 
analysis (no singularities, no ad hoc prescriptions for the 
``tangential velocities'' or for the metric), in addition to 
the numerical coincidences mentioned at the beginning of the 
section, provides some hope for pursuing a much more detailed study 
along this direction.

\section{Conclusions}
The galactic rotation curves continue to pose a challenge to present day 
physics as one would want to understand not only the nature of the 
dark matter that is associated with them but also the reason behind 
their universality (i.e., why is it distributed within a galaxy in a way 
that leads to almost flat rotation curves ?, and why is the amount of 
dark matter present in a galaxy so well correlated with the luminous 
matter ? \cite{persic,tully}). 

Models based in ordinary physical objects could already be facing problems 
(depending on the exact value of the Hubble constant \cite{copi}) 
in view of the bounds that big bang nucleosynthesis impose 
on the baryon content of the universe.

Models based on particle physics are the most commonly considered 
(usually within a Newtonian scheme) but they need to address the 
nature and the distribution problems separately, leading to a larger 
number of hypothesis and surprinsing coincidences \cite{sellwood}.

In view of the recent cosmological measurements and the theories that 
have been put forward to explain them \cite{5essence}, one is naturally 
lead to consider alternative models based in the introduction of long 
range coherent fields \cite{peebles}. In this work, we have given a 
review of various types of approaches to these questions indicating 
in each case the problems and advantages. 

We have argued that so far the most promising and simple approach would 
involve global monopoles with some sort of nonminimal coupling to 
gravity. This remains for the future to establish how far can this 
sort of ideas be pushed towards the goal of making a realistic and 
compelling model for the dynamics and evolution of galaxies. 
In particular any such model must also be studied in the context of
cosmological perturbations, large scale structure and the CMB. In this
regard we should point out that the simplest models of topological defects
as seeds for structure formation seem to be incompatible with
the acoustic peak in the CMB anisotropies detected by Boomerang and
Maxima \cite{boommax}. 
However, all these studies have considered the simplest
minimally coupled models and it is unclear how would the models of the
type being analyzed here behave in this respect. Finally, we should
mention that the currently favored cosmological scenarios require at
least two hypothetical components: the Cold Dark Matter (usually in the
form of WIMPS) necessary for the structure growth and the dark matter in
galaxies and clusters, and the cosmological constant $\Lambda$  which
provides the closure density (as required by inflation) as well as the
repulsive component that seems to be required in order to account for 
the observations of the luminosity-distance of high red shift (type Ia)
supernovae \cite{snI}. The fact that the non-minimally coupled monopoles
exhibit both an attractive regime at short distances and a repulsive
regime at large distances leads us to speculate whether these type of
models can be used to explain the two aspects of the unobserved energy
content of the universe in terms of a single hypothetical component.
Needless is to say that all these aspects will require intense further
exploration, which we hope to undertake in the near future.

\vskip 1cm
{\bf Acknowledgments}

\bigskip
U.N. is supported by a CONACyT postdoctoral fellowship
grant 990490;
M.S. and D.S. acknowledge partial support from DGAPA-UNAM Project No.
IN121298 and from CONACyT Projects 32551-E and 32272-E. Authors
thank the supercomputing department of DGSCA-UNAM.

%%%%%%%%%%%%%%%%%%%%%%%%%%%%%%%%%%%%%%%%%%%%%%%%%%%%%%%%%%%%%%%%%%%%%%%%%%

% FIGURE CAPTIONS

\vskip 1cm

%CONFIGURATIONS XI=2,0.3,0

%%%%%%%%%%%%%%%%%%%%%%%%%%%%%%%%%%%%%%%%%%%%%%%%%%%%%%%%%%%%%%%%%%%%%%%%%%

\vskip 1cm

\vskip 1cm

%%%%%%%%%%%%%%%%%%%%%%%%%%%%%%%%%%%%%%%%%%%%%%%%%%%%%%%%%%%%%%%%%%%%%%%%%

% configurations of mass vs deficit angle for different $\xi$ 

\vskip 1cm

\vskip 1cm

%%%%%%%%%%%%%%%%%%%%%%%%%%%%%%%%%%%%%%%%%%%%%%%%%%%%%%%%%%%%%%%%%%%%%%%%%%

 \vskip 1cm

 \vskip 1cm


\begin{thebibliography}{}
\bibitem{footnote}
Those contributions are: that which is 
associated to the energy density of the scalar field and the other 
which is associated to the reduction of the negative binding energy 
due to the reduction of $G$ due to a nonzero value of the scalar 
field: $G\sim 1/(1+16\pi\xi\phi^2)$ with $\xi>0$.
\end{thebibliography}
\end{document}